\documentclass[lettersize,journal]{IEEEtran}
\usepackage{xcolor}
\usepackage[table]{xcolor}
\usepackage{amsmath,amsfonts}
\usepackage{algorithmic}
\usepackage{algorithm}
\usepackage{array}
\usepackage{subcaption}
\usepackage[caption=false,font=normalsize,labelfont=sf,textfont=sf]{subfig}
\usepackage{textcomp}
\usepackage{stfloats}
\usepackage{tabularx}
\usepackage{url}
\usepackage{verbatim}
\usepackage{multirow}
\usepackage{graphicx}
\usepackage{cite}
\begin{document}

\title{A Sensor-Adaptive Incremental Learning Framework for Artifact Detection in Satellite Precipitation Data}

\author{Andres F. Monsalve$^{}$, Hernan A. Moreno,$^{}$ and Christian D. Kummerow $^{}$~\IEEEmembership{}
\thanks{A. F. Monsalve is with the University of Texas at El Paso, El Paso, TX, USA.}
\thanks{H. A. Moreno is with the University of Texas at El Paso, El Paso, TX, USA.}
\thanks{C. D. Kummerow is with Colorado State University, Fort Collins, CO, USA.}
}



\maketitle

\begin{abstract}

Historically, retrieving rainfall data from satellite imagery has been the domain of space agencies. However, in recent years, the development of cheaper, more compact satellites (SmallSats) capable of detecting rainfall proxies has led to a significant increase in private-sector initiatives for satellite launch and surface precipitation products. This rapid growth has yet to be matched by data validation efforts. Consequently, the need for a robust tool to detect anomalies in near-real-time data before it is disseminated to the public has become critical. In this paper, we present the development of an anomaly-detection system to identify artifacts in global satellite-based rainfall products. The developed framework leverages pre-trained computer vision models and incorporates scarce human-labeled data to detect specific anomalies. Our proposed anomaly detection strategy is tested on data from the Special Sensor Microwave Imager (SSMI) and the Special Sensor Microwave Imager/Sounder (SSMIS). Results demonstrate the efficacy of our approach at separating regular orbits from artifact-containing orbits for each satellite, with performance comparable to state-of-the-art in-place methods. Additionally, the framework offers explainability and the capacity for iterative refinement following false-positive or false-negative classifications.
\end{abstract}


\section{Introduction}

\IEEEPARstart{T}{he} state-of-the-art in global precipitation measurement and monitoring relies heavily on Passive Microwave (PMW) Radiometer measurements \cite{levizzaniSatellitePrecipitationMeasurement2020}. The advantages of microwave measurements over other methods, such as infrared, for measuring precipitation across large and global scales include the ability to "see through" the upper layer of clouds and the capacity to operate consistently during both day and night \cite{kiddStatusSatellitePrecipitation2011}. 

Historically, the demand for globally distributed precipitation information has been primarily supplied by government entities and alliances among multiple countries. The most important example of this type of alliance is the Global Precipitation Measurement (GPM) satellite mission. The National Aeronautics and Space Administration (NASA) of the United States, along with the Japan Aerospace Exploration Agency (JAXA), provides the "core" satellite. Together with an array of other governmental agencies (e.g., NOAA, EUMETSAT, or ISRO for India), the full constellation of satellites provides PMW measurements of Brightness Temperature ($T_B$) across the entire Earth. This constellation is able to collectively achieve an average revisit time of approximately 90 minutes for any given location \cite{skofronick-jacksonGlobalPrecipitationMeasurement2018, nasaGPMGPROFAlgorithm2022}.

The composition of ever-evolving constellations of orbiting PMW radiometers is changing even more rapidly with the advent of the CubeSat technology \cite{handCubeSatsPromiseFill2015, poghosyanCubeSatEvolutionAnalyzing2017} that allows the constellations of satellites to be built and launched through standardized processes at a much lower cost. Lego-like assembly and launch in non-exclusive rockets have made it easier \cite{karlalantElonMuskLaunching2017} to fill gaps in weather and precipitation data.

Decentralization of satellite-based radiometric data significantly increases the availability of global precipitation observations. However, unlike missions such as the GPM constellation, which benefit from rigorous quality control protocols, CubeSat missions often lack comparable oversight. While these platforms incorporate Quality Assessment (QA) procedures for Brightness Temperature ($T_B$), even minor, undetected anomalies in $T_B$ measurements can propagate into significant errors in the calculated precipitation fields \cite{strakaCloudPrecipitationMicrophysics2009,kiddPrecipitationEstimationNASA2022}.

Surface precipitation estimates from passive microwave satellites are typically produced by retrieval schemes that compare measured $T_B$ values with prior databases of matching $T_B$s for each simulated rain profile \cite{levizzaniSatellitePrecipitationMeasurement2020}. These methods are highly non-linear and thus require well-calibrated, stable data inputs. Current QA strategies generally rely on fixed physical thresholds, such as temperature limits, and on instrument health metrics such as voltage stability or orbital maneuvers. Nevertheless, these QA processes remain predominantly pixel-based and are made in stages prior to the final rain product \cite{kiddPrecipitationEstimationNASA2022}. Several in-place processes attempt to identify imperfections within precipitation retrievals, most notably the recently developed Satellite Precipitation Estimate Error Detector (SPEEDe) \cite{tanAutomatedQualityControl2024}. SPEEDe utilizes a convolutional autoencoder trained exclusively on nominal, artifact-free precipitation fields. When evaluating a new orbit, the algorithm attempts to reconstruct the data; unphysical or anomalous inputs result in high reconstruction errors, generating a score to flag the orbit. While methods like SPEEDe and other unsupervised anomaly detectors \cite{samuelUnsupervisedAnomalyDetection2021} demonstrate high proficiency in distinguishing between artifact-contaminated and regular orbits, they fundamentally lack interpretability. They provide overall scores or binary indicators, but do not identify the underlying characteristics of the anomaly. Furthermore, these methods often fail to spatially localize suspected errors within an image and remain highly limited when adapting to novel, unexpected error patterns as they emerge.

\begin{figure*}
    \centering
    \includegraphics[width=\textwidth]{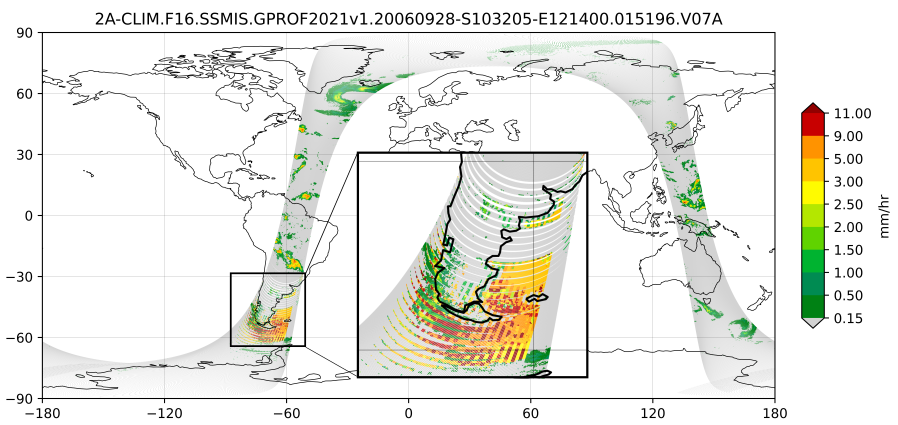}
    \caption{Identification of a precipitation retrieval artifact within SSMIS orbit \#015196 (September 28, 2006). The magnified detail window displays a specific anomaly encountered off the coasts of Chile and Argentina. The rest of the orbit is considered to have regular precipitation patterns.}
    \label{fig:1}
\end{figure*}

To illustrate the nature of these anomalies, we present an artifact identified within a Special Sensor Microwave Imager/Sounder (SSMIS) orbit (Figure \ref{fig:1}), localized off the southern coasts of Argentina and Chile. Specifically, the retrieved precipitation field in this region exhibits highly structured, parallel, and intermittent patterns of mid-to-high intensities that lack a physical meteorological basis. While the individual pixel intensities fall within a realistic range, their rigid geometric arrangement indicates a sensor or retrieval error rather than a natural precipitation system. While artifacts have been documented across several traditional PMW sensors, including SSMIS (onboard F16, F17, and F18), the MHS (onboard Metop-B and Metop-C), the ATMS (onboard NOAA-19 and NOAA-20), and the legacy Special Sensor Microwave Imager (SSMI; onboard F08), the total number of recorded instances remains limited due to the inherent reliability of the sensors and the highly labor-intensive nature of manually identifying subtle artifact signatures.

Despite these constraints, the available examples curated by personnel from the NASA Precipitation Processing System (PPS) provide a robust foundation for developing an incremental learning framework to adaptively identify artifacts in surface precipitation fields, validated against SSMIS and SSMI data. While the current inventory of confirmed anomalies in CubeSat-based missions is relatively small, the framework proposed here is specifically designed to adapt to these smaller satellites. Legacy sensors such as SSMI and SSMIS utilize conical scanning mechanisms that maintain a constant field of view (FOV), whereas current CubeSat radiometers primarily employ cross-track scanning \cite{blackwellOverviewTROPICSNASA2018, brownSinglePointCalibrationMicrowave2023}, introducing FOV elongation at high scan angles. Despite this geometric divergence, the proposed patch-based convolutional approach remains applicable. By establishing a baseline through transfer learning and subsequently ingesting sensor-specific data, the network implicitly learns the spatial biases of the target instrument (such as swath edges FOV distortions). This allows the framework to dynamically adapt to the morphological transformations characteristic of individual sensor operations, rather than relying on static spatial assumptions.

The specific causes of the artifacts presented in this research remain unidentified; they could be attributed to atmospheric and surface-type conditions, such as layers of ice in the atmosphere or on the ocean surface, or emissivity transitions between varying surface types, as well as to mechanical causes related to the instrument operation \cite{kiddPrecipitationEstimationNASA2022}. Determining the precise cause of each artifact would require extensive independent verification and is beyond the scope of this work. Instead, this research focuses on the automated identification of artifacts in new data outputs in near real time. By establishing a robust detection framework, we aim to enhance the reliability of PMW global precipitation products and facilitate the broader operational adoption of CubeSat-based constellations. To achieve this, we introduce a sensor-adaptable, patch-based computer vision framework. Specifically, the proposed approach utilizes a deep convolutional neural network initialized via transfer learning, which is incrementally updated using human-labeled image patches to spatially isolate and classify morphological anomalies within the retrieved precipitation fields.

In chapter \ref{chap:data} of this article, we describe the data used to support our hypothesis. In chapter \ref{chap:methodology}, we provide a step-by-step description of how the framework operates jointly for two instruments across different time frames and types of artifacts affecting them. In chapter \ref{chap:results}, we present the results and discuss them more in-depth in chapter \ref{chap:discussion}, and finally conclude about the significance of this article for the ever-growing availability of surface precipitation data in chapter \ref{chap:conclusion}.

\section{Data}\label{chap:data}

This study utilized data from two microwave radiometer instruments: the Special Sensor Microwave Imager (SSMI) and the Special Sensor Microwave Imager/Sounder (SSMIS). For the SSMI instrument, data were sourced exclusively from the Defense Meteorological Satellite Program (DMSP) F-8 \cite{GPM2022ssmiF08} satellite. For the SSMIS instrument, data from the DMSP F-16 \cite{GPM2022ssmisF16}, F-17 \cite{GPM2022ssmisF17}, and F-18 \cite{GPM2022ssmisF18}  satellites were used \cite{SSMISgrumman2002adum}.

The GPROF V7 datasets used by the GPM project to create SSMI and SSMIS precipitation datasets were acquired from the NASA Goddard Earth Sciences Data and Information Services Center (GES DISC). To obtain orbits with artifacts for the purposes of this study, supplementary data were provided by personnel from the PPS, who manually identified, curated, and archived these anomalous cases. The curated dataset comprising the complete set of labeled samples for both artifact and regular classes, for SSMI and SSMIS, is publicly available as a citable data deposit on Zenodo (\url{https://doi.org/10.5281/zenodo.21793955}). This deposit contains all
processed patches used for model training and validation prior to augmentation;
the augmented samples are not archived separately, as they are deterministically
reproducible from the published patches through the augmentation protocol detailed in Section \ref{step:7} and the accompanying code
(\url{https://github.com/afmonsalves/SAPAD}).

The DMSP satellites operate in a sun-synchronous, polar orbit. The SSMI instrument series was operational from 1987 to 2008; for this study, the record for the period 1987-1991 was used. The SSMIS instrument series began operations in 2000 and continues to date; however, due to the high data volume, this study includes randomly sampled orbits from 2006 to 2024. A complete list of the orbits used in this analysis is provided in the \texttt{orbit\_lists/} folder of
the accompanying code repository (\url{https://github.com/afmonsalves/SAPAD.git}).

SSMI and SSMIS instruments measure microwave radiances from the Earth's surface by rotating a parabolic reflector counterclockwise at 31.6 rpm, obtaining a new row of data every 2 seconds within the 143.2° active angle of the radar. The satellite moves forward while performing this movement, allowing for a 12.5 km spacing in both cross-track and along-track directions. The process results in 14 new orbits every day \cite{fennigFundamentalClimateData2022}. Measured radiance values are transformed to surface precipitation values through the Goddard Profiling Algorithm (GPROF), which succinctly speaking, is a Bayesian algorithm that relates previously simulated Brightness Temperatures ($T_B$) to their corresponding precipitation profile, and then compares the simulations against measured $T_B$ values to give the final prediction of surface precipitation. To see details on the GPROF algorithm, refer to \cite{kummerowEvolutionGoddardProfiling2015}.

The fundamental unit of data in this study is the orbit, which is stored as a single file in either Hierarchical Data Format 5 (HDF5) or Network Common Data Form (netCDF) format, acquired from the aforementioned GES DISC archive. These files contain surface rain rate information (mm/h) alongside the corresponding latitude and longitude coordinates. The cross-track scan width is 128 pixels for SSMI and 180 pixels for SSMIS, while the along-track dimension typically comprises 3,210 and 3,220 ± 10 scans per orbit, respectively.

The emphasis panel on the lower left section of Figure \ref{fig:1} draws attention to an example of what is considered an artifact in the context of this research: rainfall features whose individual pixel values lie within the range of normal values, but altogether have an odd shape. While the term "artifact" in satellite imagery can be subjective, this study employs specific definitions to ensure consistent identification. We define six distinct classes of artifacts based on known retrieval errors from SSMI (1987–1991) or SSMIS (2006–2025) sensors: Lines, Spots, Bands, Landmask, Smooth, and Mosaic. Table \ref{table:1} provides detailed descriptions and examples for each class.

\begin{table*}[!p] 
    \centering
    \caption{Image dictionary with example patches of artifacts discovered for SSMI or SSMIS orbits.}
    \label{table:1}

    \renewcommand{\arraystretch}{3.8}

    \begin{tabularx}{\textwidth}{| >{\raggedright\arraybackslash}m{0.25\textwidth} | >{\centering\arraybackslash}X |}
        \hline
        {\Large \textbf{Lines}} \newline \newline \small Structured artifacts appearing in a cross-track orientation. Composed of single-row lines moderate to low intensities. \newline \newline \textbf{Present in}: SSMI, SSMIS& 
        \begin{tabular}{ccc}
            \includegraphics[width=0.27\linewidth]{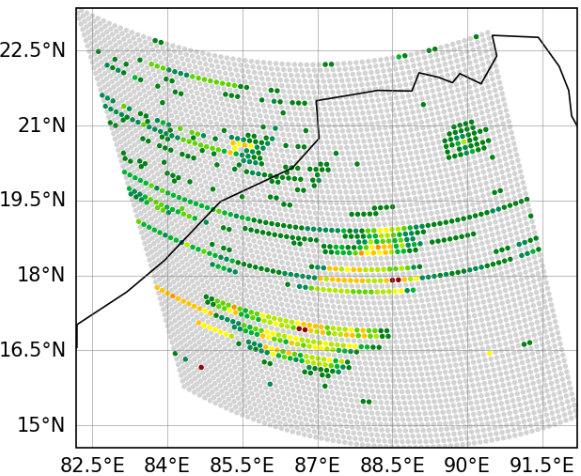} & \includegraphics[width=0.27\linewidth]{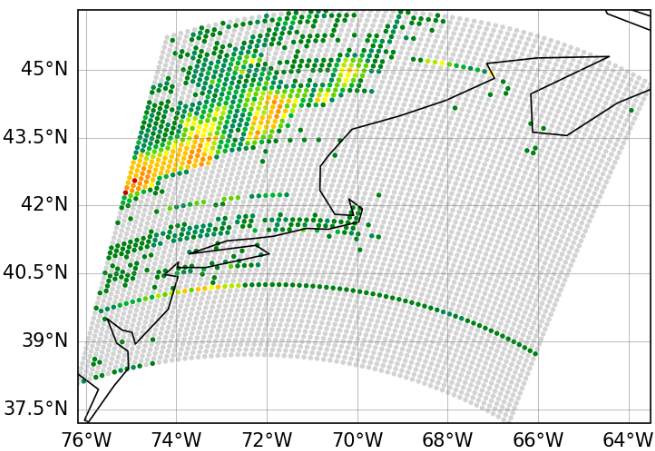} &
            \includegraphics[width=0.27\linewidth]{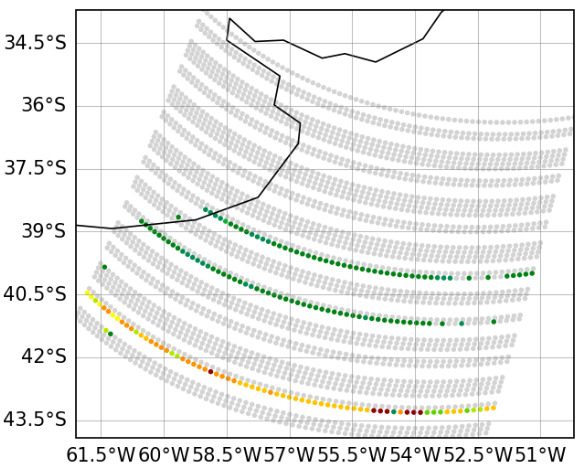}\\[-6ex]
            SSMI, orbit 004446 & SSMI, orbit 006154  & SSMIS, orbit 014899
        \end{tabular} \\
        \hline

        {\Large \textbf{Spots}} \newline \newline \small Unstructured and scattered artifacts, composed of single pixel spots, frequently with low intensities. \newline \newline \textbf{Present in}: SSMI, SSMIS & 
        \begin{tabular}{ccc}
            \includegraphics[width=0.27\linewidth]{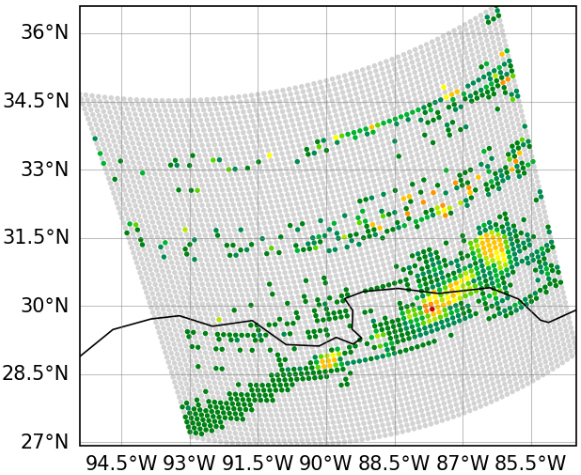} & \includegraphics[width=0.27\linewidth]{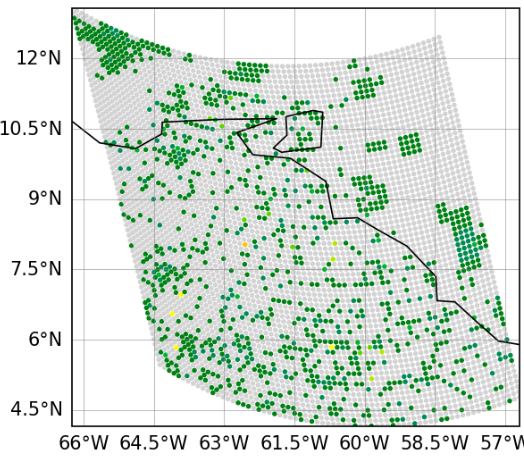} &
            \includegraphics[width=0.27\linewidth]{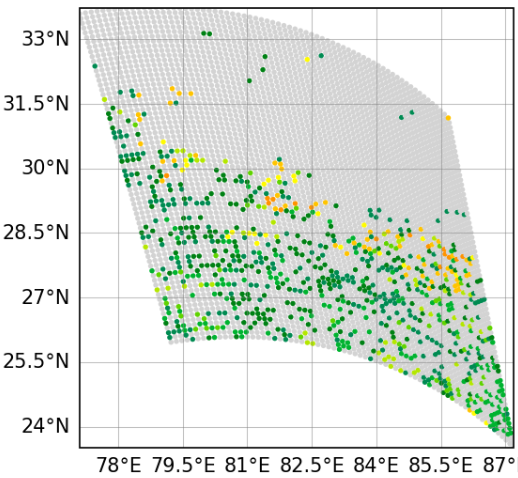}\\[-6ex]
            SSMI, orbit 004114 & SSMI, orbit 007813  & SSMIS, orbit 01198
        \end{tabular} \\
        \hline

        {\Large \textbf{Bands}} \newline \newline \small Structured parallel along-track artifacts. Composed of high and low intensities. \newline \newline \textbf{Present in}: SSMIS & 
        \begin{tabular}{ccc}
            \includegraphics[width=0.3\linewidth]{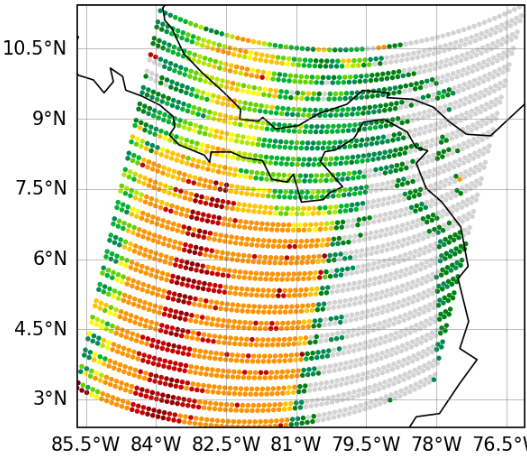} & \includegraphics[width=0.3\linewidth]{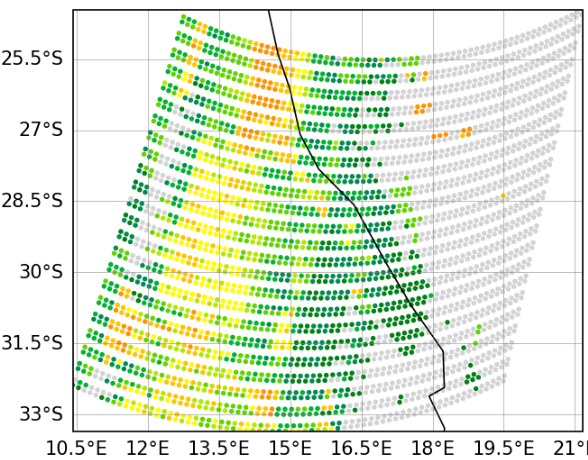}& \includegraphics[width=0.3\linewidth]{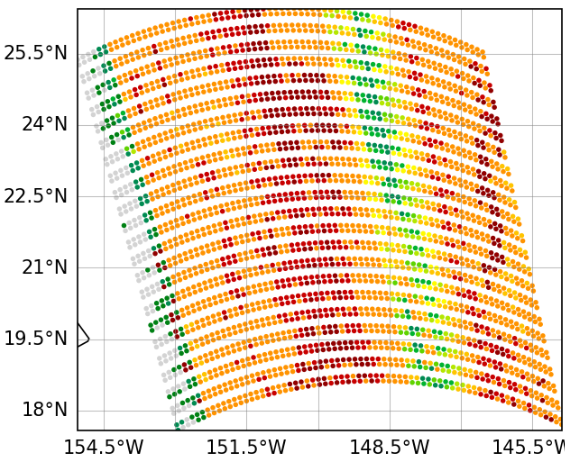}\\[-6ex]
            SSMIS, orbit 067944 & SSMIS, orbit 067954 & SSMIS 067954
        \end{tabular} \\
        \hline

        {\large \textbf{Landmask}} \newline\newline \small Structured artifacts where the rain features follow the coastline with varying intensities. \newline \newline \textbf{Present in}: SSMIS & 
        \begin{tabular}{ccc}
            \includegraphics[width=0.27\linewidth]{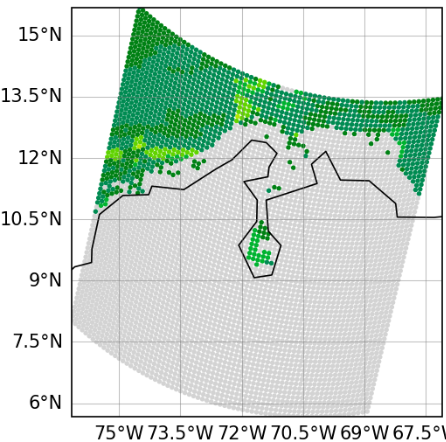} & \includegraphics[width=0.27\linewidth]{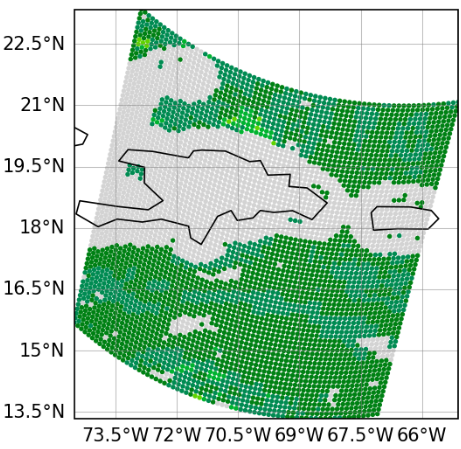} &
            \includegraphics[width=0.27\linewidth]{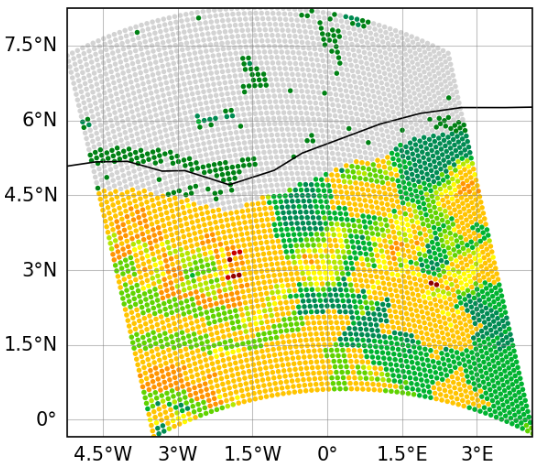}\\[-6ex]
            SSMIS, orbit 045162 & SSMIS, orbit 045162 & SSMIS, orbit 095608
        \end{tabular} \\
        \hline
        
        {\Large \textbf{Smoothness} } \newline\newline \small Rain features with abnormally smooth geometrical shapes and sharp boundaries. \newline \newline \textbf{Present in}: SSMIS & 
        \begin{tabular}{ccc}
            \includegraphics[width=0.25\linewidth]{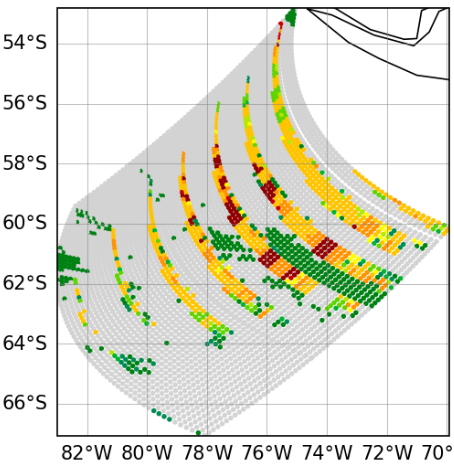} & \includegraphics[width=0.27\linewidth]{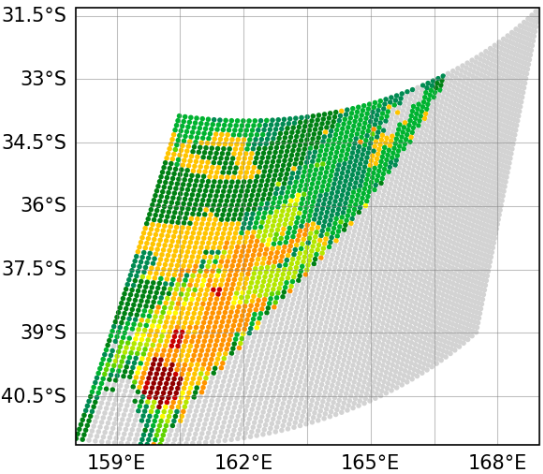} &
            \includegraphics[width=0.3\linewidth]{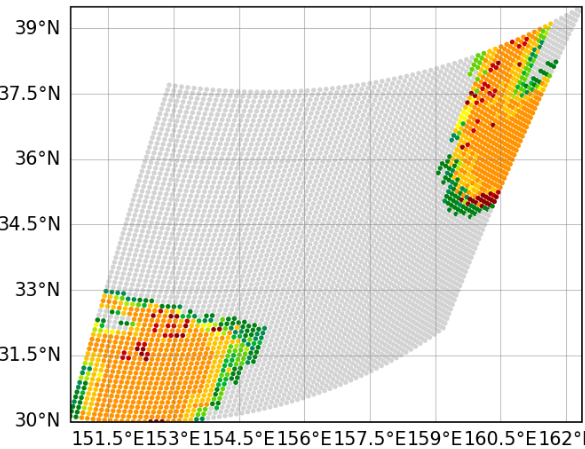}\\[-6ex]
            SSMIS, orbit 014899 & SSMIS, orbit 075475  & SSMIS, orbit 075476
        \end{tabular} \\
        \hline

        {\Large \textbf{Mosaic} } \newline\newline \small Discontinuous features that resemble mosaic characteristics. \newline \newline \textbf{Present in}: SSMIS & 
         \begin{tabular}{ccc}
            \includegraphics[width=0.28\linewidth]{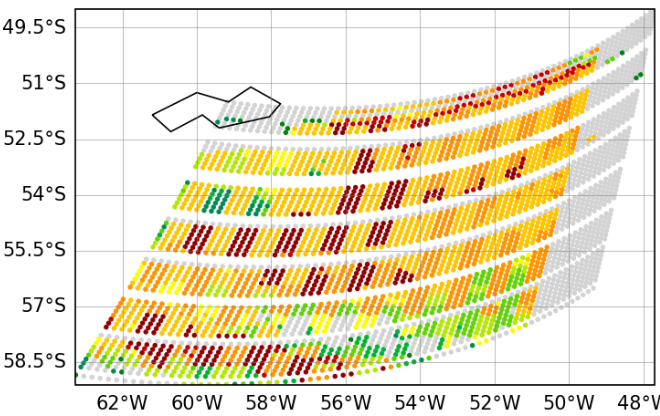} & \includegraphics[width=0.28\linewidth]{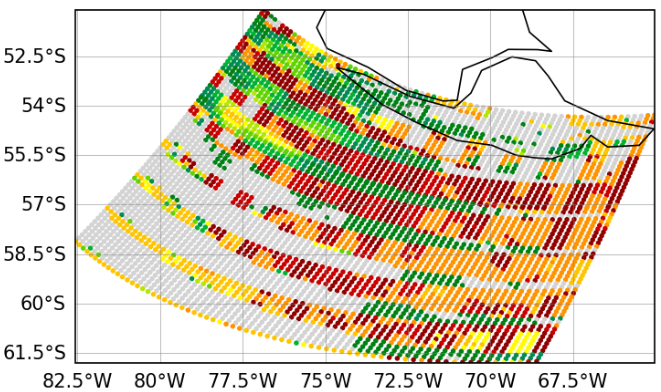} & \includegraphics[width=0.28\linewidth]{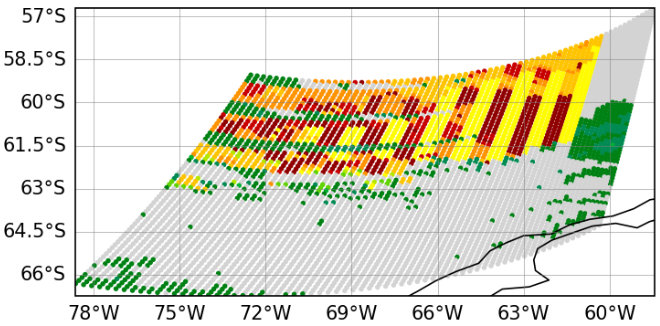}\\[-6ex]
            SSMIS, orbit 014899 & SSMIS, orbit 015196 & SSMIS, orbit 015196
        \end{tabular} \\
        \hline
    \end{tabularx}
\end{table*}

The occurrence of orbits containing artifacts is statistically infrequent relative to regular observations; this scarcity results in a heavily imbalanced dataset. The distribution of total orbits used in this study, categorized as regular orbits or those containing identified anomalies, is summarized in Table \ref{tab:combined_anomaly_count}.

\begin{table}[H]
    \centering
    \caption{Counts of SSMI and SSMIS orbits exhibiting either no artifacts, specific artifact types, or combinations of artifacts (see Table \ref{table:1}), spanning 1987–1991 for SSMI and 2006–2024 for SSMIS.}
    \label{tab:combined_anomaly_count}
    \begin{tabular}{|p{4.2cm}|c|c|}
        \hline
        \textbf{Artifact Type} & \textbf{SSMI} & \textbf{SSMIS} \\
        \hline
        Regular (no artifacts) & 167 & 193 \\
        Lines only & 7 & 0 \\
        Spots only & 11 & 2 \\
        Bands only & 0 & 7 \\
        Landmask only & 0 & 1 \\
        Smoothness only & 0 & 0 \\
        Mosaic only & 0 & 0 \\
        Multiple artifacts (any combination) & 2 & 9 \\
        \hline
    \end{tabular}
\end{table}

Although the absolute count of anomalous orbits is small, the spatial decomposition of these orbits into overlapping sub-windows significantly amplifies the training data volume. By employing a sliding-window extraction protocol (detailed in Section \ref{step:2}), a single anomalous orbit yields tens of highly localized positive samples. This transformation along with data augmentation, effectively converts a sparse orbit-level dataset into a dense patch-level dataset, generating the spatial variance necessary to robustly train Deep Convolutional Neural Networks.

\section{Methodology}\label{chap:methodology}

The operational conditions of passive microwave radiometers are not universal, leading to often specific issues and problems. To overcome the challenge of creating a universal artifact-identification process that accommodates these differences, this study employs a patch-based framework. In this context, a "patch" refers to a discrete sub-window that slides across the domain of any given satellite product. This approach allows the framework to operate independently of the specific output details of precipitation products, making it effectively sensor-agnostic.

A significant motivation for this methodology is the limitation of existing automated quality controls. Many of the rainfall artifacts identified in this study occur in addition to the pixels flagged in the standard Level 1C quality control procedures.   \cite{precipitationprocessingsystemppsAlgorithmTheoreticalBasis2016}. Because these anomalies are often only detectable through subsequent manual review, this study utilizes human-labeled orbits as the "ground truth" for training and validation, aiming to automate a detection process that currently requires intensive manual oversight. 

All processing made to data, including ingestion, labeling, processing, and performance assessments were conducted using Python-based scripts in a Jupyter Notebook environment \cite{grangerJupyterThinkingStorytelling2021}. The hardware used was a Dell Precision Workstation 7920 equipped with 64GB of RAM and 16 processing cores. The computational model training, was conducted within the Google Colaboratory cloud computing environment (\textit{\url{https://colab.research.google.com/}}), utilizing an NVIDIA T4 Tensor Core GPU. The methodology is detailed through ten distinct components that define the framework’s design. 

\subsection{Data Visualization}\label{step:1}

In this framework, data visualization is a critical diagnostic step. Because artifact identification relies on meteorologists' experience at recognizing regular and odd-looking patterns, it is imperative to maintain a consistent visual standard that enables reliable detection of spatial anomalies. To ensure this consistency, all geolocated precipitation pixels were plotted using the Cartopy Python library \cite{Cartopy} under a Plate Carrée map projection. Furthermore, to eliminate variability in human interpretation, a standardized colorbar and fixed value limits, detailed in Table \ref{table:colors}, were applied across all orbits. This standardization ensures that specific rainfall features, such as sustained intensities and geometric patterns, remain visually comparable regardless of the specific sensor or orbit being analyzed.

\begin{table}[h]
    \centering

    \caption{Colormap used for surface precipitation visualization across all study images.}
    \label{table:colors}
    \renewcommand{\arraystretch}{1.2} 
    \begin{tabular}{|l| c| l| l|}
        \hline
        \textbf{Range (mm/h)} & \textbf{Color} & \textbf{Hex Code} & \textbf{RGB (0-255)} \\
        \hline
        $<$ 0.15 & \cellcolor[HTML]{3A3D48} & \texttt{\#3A3D48} & (58, 61, 72) \\
        0.15 - 0.50 & \cellcolor[HTML]{008114} & \texttt{\#008114} & (0, 129, 20) \\
        0.50 - 1.00 & \cellcolor[HTML]{008B52} & \texttt{\#008B52} & (0, 139, 82) \\
        1.00 - 1.50 & \cellcolor[HTML]{00B330} & \texttt{\#00B330} & (0, 179, 48) \\
        1.50 - 2.00 & \cellcolor[HTML]{60D300} & \texttt{\#60D300} & (96, 211, 0) \\
        2.00 - 2.50 & \cellcolor[HTML]{B4E700} & \texttt{\#B4E700} & (180, 231, 0)\\
        2.50 - 3.00 & \cellcolor[HTML]{FFFB00} & \texttt{\#FFFB00} & (255, 251, 0)\\
        3.00 - 5.00 & \cellcolor[HTML]{FFC400} & \texttt{\#FFC400} & (255, 196, 0)\\
        5.00 - 9.00 & \cellcolor[HTML]{FF9300} & \texttt{\#FF9300} & (255, 147, 0)\\
        9.00 - 11.00 & \cellcolor[HTML]{FF0000} & \texttt{\#FF0000} & (255, 0, 0) \\
        $>$ 11.00 & \cellcolor[HTML]{c80000} & \texttt{\#C80000} & (200, 0, 0) \\
        \hline
    \end{tabular}
\end{table}

\subsection{Patch-based Data Preparation Pipeline}\label{step:2}

\begin{figure}[h!]
    \centering
    \includegraphics[width=\linewidth]{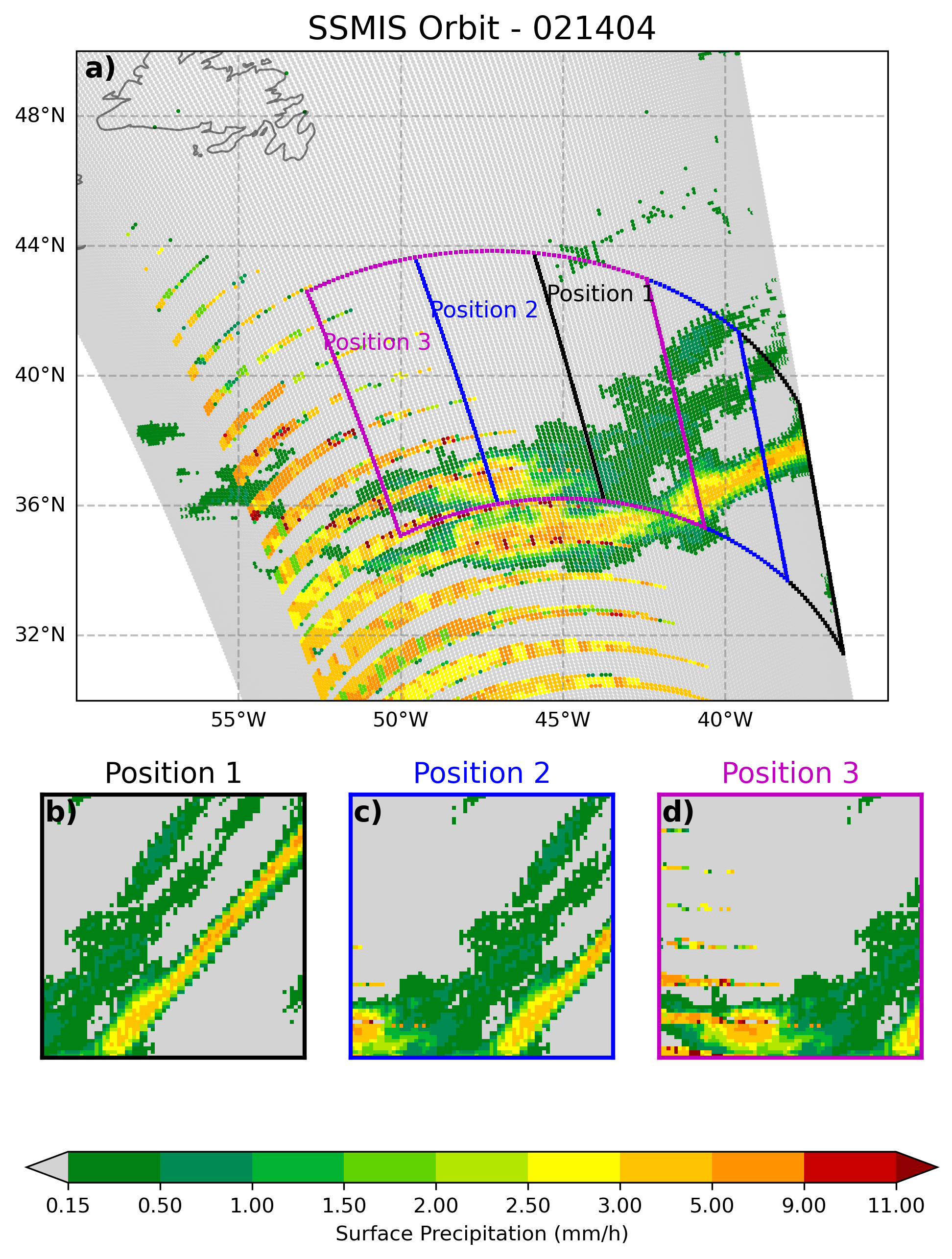}
    \caption{Schematic of the sliding-window patch extraction through each orbit and manual auditing process. (a) Representative satellite orbit section illustrating the kernel traversal path. (b, c) Examples of candidate window positions discarded during the manual audit due to the absence of clear artifact signatures or the presence of ambiguous meteorological features. (d) A confirmed positive patch containing a distinct 'Smooth' artifact, subsequently included in the training pool. The overlapping grid structure (23-pixel stride) ensures that artifact features are fully captured within the $70 \times 70$ pixel window boundaries.}
    \label{fig:positions}
\end{figure}

To isolate specific artifact features within an orbit (e.g., Orbit \#015196 in Fig. \ref{fig:1}), we implemented a sliding-window extraction method. As illustrated in Figure \ref{fig:positions}, a kernel of $70 \times 70$ pixels traverses the orbit domain. This specific window dimension was selected through empirical optimization: it provides a spatial extent large enough to capture the morphological nuances of individual artifacts, yet small enough to effectively increase the training sample volume through sub-windowing. Given the nominal 12.5 km pixel spacing for both the SSMI and SSMIS Level 2 precipitation products, a  $70 \times 70$ pixel window corresponds to a footprint of approximately 875 km by 875 km on the surface. To ensure the highest possible density of training features and to prevent loss of artifacts at the window boundaries, a stride of one-third of the window dimensions (23 pixels) was applied in both the cross-track and along-track directions. This results in a $66\%$ spatial overlap between adjacent positive patches. The images were manually audited at each window position, and only patches containing an unambiguous and clearly defined artifact signature were included in the positive training pool (Fig. \ref{fig:positions}, panel \textbf{d}). Windows that appeared devoid of artifact features but originated from an anomalous orbit (Fig. \ref{fig:positions}, panel \textbf{b}) were completely excluded from all training data to prevent potential leakage of spatial correlations. Crucially, windows capturing only marginal or subtle traces of an artifact at the boundaries (as in the lower-left corner of Fig. \ref{fig:positions}, panel \textbf{c}) were assigned to an auxiliary 'IDK' category (which is discussed in detail in Section \ref{chap:discussion}). Injecting these ambiguous, truncated signals into the formal training classes would introduce label noise, which could degrade the model's convergence and discriminative capabilities.

Negative samples, or typical rainfall features, were extracted using a different sampling logic. Regular orbits were partitioned into $70 \times 70$ pixel patches using a non-overlapping grid. The process over regular orbits naturally yields a high frequency of "no-rain" patches. To prevent the model from developing a bias toward the trivial case of empty data, an undersampling strategy was applied such that only $20\%$ of the total identified "no-rain" patches were retained for the final training and validation datasets, ensuring the framework remains sensitive to the more complex features of both legitimate precipitation and artifacts.

\subsection{Model Baseline Architecture Selection}\label{step:3}

For this study, Convolutional Neural Network (CNN) architectures were considered over more recent alternatives, such as Vision Transformers (ViT), because it has been tested that the inductive bias of CNNs helps them avoid overfitting for smaller datasets as compared to ViTs \cite{dosovitskiyImageWorth16x162021}. The modeling pipeline used the PyTorch library \cite{anselPyTorch2Faster2024}, which provides an extensive collection of pre-trained Deep Convolutional Neural Network (DCNN) architectures. 

The optimal balance between model performance and computational overhead was achieved with the MobileNetV3-Large architecture, which was chosen as the primary architecture. The selection was based on a series of experiments across different model families. During initial testing, MobileNetV3 Small was discarded due to insufficient capabilities, resulting in lower detection accuracy. Residual networks were tested as well \cite{heDeepResidualLearning2015}; while ResNet-18 achieved marginally better performance metrics than MobileNetV3-Large, it required significantly longer training. Paradoxically, larger architectures such as ResNet-50 and ResNet-101 yielded inferior results; we hypothesize that these higher-parameter models suffered from overparameterization leading to "data hunger" and subsequent overfitting. This hypothesis was not further tested in this study.

\subsection{Hyperparameter Tuning}\label{step:5}

An open software optimization framework, Optuna \cite{akibaOptunaNextgenerationHyperparameter2019}, was used to fine-tune training hyperparameters such as the optimizer, batch size, initial learning rate, and weight decay. 

\subsection{Multi-Channel Input Strategy}\label{step:6}

To maintain compatibility with the three-channel ($RGB$) input requirements of the pre-trained MobileNetV3 Large architecture, each $70 \times 70$ pixel precipitation patch is restructured into a multi-channel tensor. The first channel is populated with precipitation values, scaled globally to 0-11 $mm/hr$, providing the model with a consistent absolute reference for rainfall intensity across all orbits. The second channel uses local scaling, where values are normalized between 0 and 1; this enhances the model’s sensitivity to local subtle artifact textures that might be obscured by global scaling. Finally, the third channel consists of a binary landmask that delineates land and ocean surfaces within each sub-window. This input was incorporated to provide explicit spatial context intended to aid the network in learning the distinctive coastal geometries of the 'Landmask' artifact class.

\subsection{Training Data Augmentation}\label{step:7}

A data augmentation pipeline was implemented to artificially expand the training dataset through  image transformations. Each 70×70-pixel patch underwent three primary transformations: reflection, translation, and rotation. Specifically, patches are reflected along both the horizontal (left-right) and vertical (up-down) axes. Spatial translations are achieved via circular shifts, or "rolls," where the data is shifted by 15 pixels in both the vertical and horizontal directions. Furthermore, rotation is introduced by generating variants at 90°, 180°, and 270° increments.

These specific, deterministic transformations were selected over random continuous augmentations (such as arbitrary angle rotations, shearing, or scaling) to maintain strict control over the morphological integrity of the synthetic samples. Because the framework must distinguish distinct artifact topologies, continuous transformations requiring pixel interpolation were strictly avoided. Interpolation mathematically alters raw precipitation intensity values and blurs sharp boundary gradients, which risks generating synthetic features that mimic different artifact classes, thereby introducing label confusion. The 90-degree rotational increments and orthogonal reflections represent rigid transformations that perfectly preserve the original pixel intensities and exact geometric relationships of the anomalies. The 15-pixel spatial translation (approximately 21\% of the window domain) was selected because it provides sufficient spatial displacement to build translation invariance within the network, while remaining smaller than the 23-pixel sliding window extraction stride. The selected transformations ensure that the augmented dataset represents spatial variations without corrupting the unique morphological signature of the target artifacts.

\subsection{Model Training and Validation}\label{step:4}

\begin{figure}[htbp]\captionsetup[subfigure]{font=footnotesize}
    \centering
    \begin{subfigure}{0.35\linewidth}
        \centering
        \includegraphics[width=\linewidth]{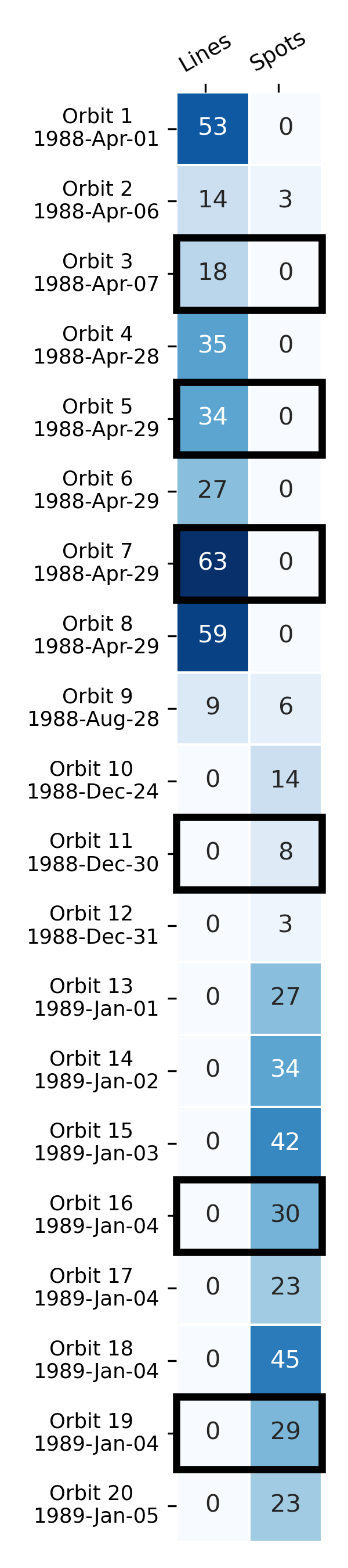}
        \caption{SSMI orbits with artifacts.}
        \label{fig:left}
    \end{subfigure}
    \hfill 
    \begin{subfigure}{0.63\linewidth}
        \centering
        \includegraphics[width=\linewidth]{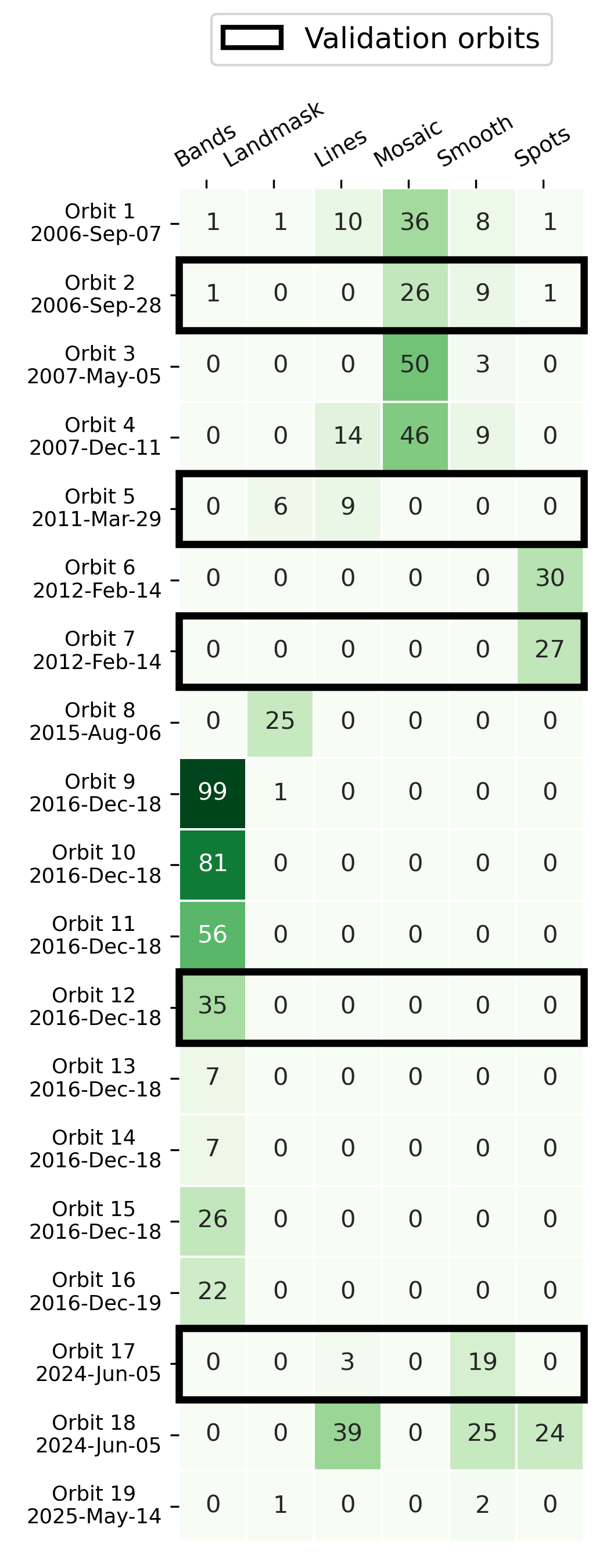}
        \caption{SSMIS orbits with artifacts.}
        \label{fig:right}
    \end{subfigure}
    
    \caption{Available orbital scans containing artifacts for (a) SSMI and (b) SSMIS. Numbers within each box indicate the total number of samples corresponding to each artifact category prior to data augmentation. Black-bordered boxes denote the subset of data reserved for validation.}
    \label{fig:Heatmap_cycles}
\end{figure}

We developed an experimental setup to evaluate the framework's ability to improve at detecting artifacts starting from a state of zero prior exposure to specific artifact classes. As new categories and artifact examples appear, the models must demonstrate gradual improvement in detection performance. The experimental data, illustrated in Figure \ref{fig:Heatmap_cycles}, consist of available orbits for SSMI and SSMIS sensors, ordered by their real order of appearance. The training progresses in sequential "cycles," where each cycle introduces a new batch corresponding to a single orbit containing arbitrary artifact classes. Following each training cycle, model performance is evaluated against a fixed hold-out validation set comprising six orbits per satellite. To explicitly detail the resulting proportion of training to validation data, Table \ref{tab:data_split} summarizes the absolute counts of original and augmented patches utilized by the final experimental cycle. The data augmentation protocol (detailed in Section \ref{step:7}) was applied exclusively to the training samples extracted from the non-validation orbits presented in Figure \ref{fig:Heatmap_cycles}, ensuring the validation set remained strictly unaugmented.

\begin{table*}[t]
\centering
\caption{Cumulative patch distribution for training and validation datasets utilized in the final experimental stage (Cycle 15). The dataset partition is based on chronological orbit separation rather than random extraction. Data augmentation was strictly isolated to the training subsets.}
\label{tab:data_split}
\begin{tabular}{llcccc}
\hline
\textbf{Sensor} & \textbf{Model Strategy} & \textbf{Original Training Patches\textsuperscript{a}} & \textbf{Augmented Patches} & \textbf{Total Training Patches} & \textbf{Validation Patches (Hold-out)\textsuperscript{b}} \\ \hline
\multirow{3}{*}{SSMI} & Baseline & 5,088 & 3,162 & 8,250 & 1696 \\
 & Warm-Start & 5,088 & 3,162 & 8,250 & 1696 \\
 & Joint & 11,139\textsuperscript{c} & 3,162 & 14,301 & 1696 \\ \hline
\multirow{3}{*}{SSMIS} & Baseline & 8,763 & 5,427 & 14,190 & 2145 \\
 & Warm-Start & 8,763 & 5,427 & 14,190 & 2145 \\
 & Joint & 12,342\textsuperscript{d} & 5,427 & 17,769 & 2145 \\ \hline
\end{tabular}
\vspace{1ex}
{\raggedright \small 
\\
\textsuperscript{a} Comprises both regular precipitation patches and unaugmented artifact patches. \\
\textsuperscript{b} Validation data include 35 standard orbits per sensor, alongside an independent set of artifact-containing orbits (6 SSMI and 5 SSMIS). \\
\textsuperscript{c} Includes an additional 6,051 cross-sensor artifact patches ingested from the SSMIS dataset. \\
\textsuperscript{d} Includes an additional 3,579 cross-sensor artifact patches ingested from the SSMI dataset. \par}
\end{table*}

This temporal design simulates the operational lifecycle of a newly deployed satellite mission. Upon initial deployment, a sensor typically transmits data assumed to be nominal. Over time,  routine scientific validation of the data identifies systematic retrieval errors, and a catalog of confirmed anomalies accumulates. By ordering the training data chronologically and introducing these anomalies in sequential cycles, the validation framework mirrors this real-world operational scenario.

Within each cycle, the training data is partitioned using a 5-fold stratified cross-validation strategy. To ensure stability when dealing with infrequent artifact classes, the number of folds is adaptively adjusted to the minimum of 5 or the lowest number of samples available for any single class up to that cycle. During the cross-validation phase, early stopping is implemented with a patience of 7 epochs, monitored by the $F_2$ score of the validation fold (see section \ref{step:9} for rationale of $F_2$ selection). To produce the definitive model for a given cycle and training strategy (see details of training strategies in section \ref{step:strategies}), the training epochs that triggered early stopping across all five folds are recorded. The final model for that strategy is then trained on the complete dataset (all folds) for a duration equal to the median value of these recorded epochs. 

Given that the definitive models for each cycle are trained for a fixed number of epochs without dynamic early stopping, we implemented a deterministic learning rate scheduler. The optimization utilizes a cosine annealing schedule to regulate the learning rate independently of real-time performance metrics. The schedule starts at a peak learning rate determined through prior hyperparameter tuning and decays to 0.01 times the initial value over 50 epochs. This ensures a controlled reduction in step size as the networks approach an optimal weight configuration for the newly introduced data.

\subsection{Cross-Sensor Data and Transfer Learning Strategies}\label{step:strategies}

A fundamental concern in applying this research is whether to train specialized models for each instrument or to deploy a single, unified model that leverages aggregate data across multiple sensors. To systematically investigate the impact of cross-sensor data inputs versus isolated learning, and to determine if larger data volumes yield superior detection models despite differing sensor geometries, three distinct training paradigms were implemented:

\begin{enumerate}
    \item \textbf{Baseline (Target-Only) Model:} This paradigm serves as the experimental control. The network is initialized with MobileNetV3's pre-trained weights and fine-tuned exclusively on the target sensor's data. The training strictly follows the incremental orbit cycles described in Section \ref{step:4}. This model represents the standard, isolated deployment scenario in which an algorithm learns its feature representations exclusively from the specific instrument it is intended to monitor, constrained by limited initial data availability.
    
    \item \textbf{Warm-Start (Sequential Transfer) Model:} This paradigm replaces the ImageNet initialization with the final, fully converged weight configuration of the opposing sensor's Baseline model (e.g., initializing an SSMIS model with final SSMI weights). During the incremental cycles, this model is exposed only to the target sensor's newly arriving data. This strategy tests whether pre-learned representations of artifacts accelerate gradient convergence or improve the absolute performance ceiling relative to the Baseline.
    
    \item \textbf{Joint (Data Fusion) Model:} Addressing the conception in Deep Learning that maximizing training volume improves generalization \cite{sun_revisiting_2017}, the Joint model aggregates data across both sensors. Initialized with MobileNetV3 weights, this network is trained during each cycle using all available artifact examples from the opposing sensor, merged with the incrementally available batches from the target sensor. To accommodate the different artifact typologies across sensors (e.g., SSMI lacking Bands or Mosaic patterns present in SSMIS), the Joint model's output layer is standardized to a six-class architecture.
\end{enumerate}

\subsection{Model Performance Assessment}\label{step:9}

All validation metrics reported were calculated using a holdout validation dataset. The dataset for each satellite selected orbits (validation orbits in Figure \ref{fig:Heatmap_cycles}) remained strictly isolated and was never presented to the algorithm during any stage of training, hyperparameter tuning, or cross-validation. 

In this study, prioritizing the detection of positive samples, i.e. artifacts, is critical; failing to detect an artifact (i.e., a false negative) is significantly more detrimental to data integrity than falsely flagging a regular orbit as an anomaly.

To focus on the positive class predictions when evaluating a model, two metrics are commonly used: Precision and Recall. The former is the ratio between the true positive predictions and all positive class predictions (true positives plus false positives). Equation \ref{eq:precision} shows the precision formula. Recall is the ratio between the true positive predictions and all positive samples (true positive plus false negative predictions). Recall formula is shown in Equation \ref{eq:recall}.

The primary evaluation metric in this study is a composite of precision and recall, called the F-score. The F-score formula is shown in Equation \ref{eq:fbeta} and has been used as a robust classification metric since 1980 \cite{rijsbergenInformationRetrieval1981}. The parameter $\beta$ in Equation \ref{eq:fbeta} allows for tuning the importance given between Recall and Precision, when $\beta$=1, the F-score becomes the harmonic mean of precision and recall. For our problem, a $\beta$=2 during evaluation is chosen because it is twice as important not to let positive classes undetected than to flag a negative sample falsely as positive.

In addition to classification accuracy, we assess the reliability of the model's probabilistic outputs using the Brier Score \cite{brierVERIFICATIONFORECASTSEXPRESSED1950} (Eq. \ref{eq:brier}). This metric measures the mean squared difference between the predicted probability ($p_i$) and the actual outcome ($o_i \in \{0, 1\}$). A lower Brier Score indicates that the model is not only more accurate but also better calibrated, meaning the assigned probabilities more closely reflect the true likelihood of an artifact's presence.

Aditionally, to evaluate the overall diagnostic capacity of the proposed framework, we benchmarked our models against the SPEEDe algorithm \cite{tanAutomatedQualityControl2024}. Because SPEEDe relies on an aggregate Mean Squared Error (MSE) rather than explicit classification probabilities, we utilized the Precision-Recall Area Under the Curve (PR-AUC) to provide a threshold-agnostic comparison of anomaly detection performance.

Given the highly imbalanced nature of the dataset, where not only artifacts are statistically infrequent relative to regular observations, but there is a big imbalance among artifact classes present, we incorporate Balanced Accuracy ($BA$) as a supplementary classification metric. This ensures that the evaluation remains unskewed and is not artificially inflated by the model's differential ability to detect different artifact classes.

\begin{equation}
    \mathrm{Precision} = \frac{\mathrm{TP}}{\mathrm{TP} + \mathrm{FP}}
    \label{eq:precision}
\end{equation}

\begin{equation}
    \mathrm{Recall} = \frac{\mathrm{TP}}{\mathrm{TP} + \mathrm{FN}}
    \label{eq:recall}
\end{equation}

\begin{equation}
    F_{\beta} = (1 + \beta^2) \cdot \frac{\mathrm{Precision} \cdot \mathrm{Recall}}{(\beta^2 \cdot \mathrm{Precision}) + \mathrm{Recall}}, \quad \beta \ge 0
    \label{eq:fbeta}
\end{equation}

\begin{equation}
\mathrm{Brier} = \frac{1}{N} \sum_{i=1}^{N} (p_i - o_i)^2
\label{eq:brier}
\end{equation}

\begin{equation}
\mathrm{BA} = \frac{1}{2} \left( \frac{\mathrm{TP}}{\mathrm{TP} + \mathrm{FN}} + \frac{\mathrm{TN}}{\mathrm{TN} + \mathrm{FP}} \right)
\label{eq:balanced_accuracy}
\end{equation}

\subsection{Assessment of Minimum Description Length}
\label{step:descriptionlength}

To empirically determine the minimum description length, defined here as the threshold of training samples required to achieve stable detection for a given artifact morphology, we conducted an isolated, class-specific training evaluation. 

For each individual artifact class within the SSMI and SSMIS datasets, an independent model was initialized.The training process was executed incrementally. The sequence began by training the model on a single positive sample of the target class ($N=1$). In each subsequent iteration, the training pool was expanded by incorporating one additional sample ($N=2, 3, \dots$) until the complete repository of available training patches for that specific class was exhausted. At every step, the active sample pool was subjected to the standard data augmentation protocols detailed in Section \ref{step:7} to preserve the framework's operational logic.

Model performance was evaluated iteratively against the chronologically held-out validation set. To precisely quantify the learning trajectory for the isolated features, the validation metrics ($F_{2}$ score and $BA$) were calculated exclusively based on the model's capacity to detect the specific target class under evaluation.

\section{Results}\label{chap:results}

\subsection{Hyperparameter Optimization}

The search space explored continuous ranges for learning rates and weight decay, alongside discrete options for optimizers and batch sizes. The resulting optimal hyperparameters, which were fixed across all respective training cycles for each sensor, are presented in Tables \ref{tab:hyperparams_SSMI} and \ref{tab:hyperparams_SSMIS}. For the Joint (Data Fusion) training paradigm, the model strictly adopts the static hyperparameter set corresponding to the primary target sensor being evaluated (e.g., the Joint model evaluating SSMIS utilizes the fixed SSMIS hyperparameters, despite ingesting auxiliary cross-sensor data).

\begin{table}[h!]
\centering
\caption{SSMI hyperparameter tuning search space and the selected optimal values (in bold).}
\label{tab:hyperparams_SSMI}
\renewcommand{\arraystretch}{1.3} 
\begin{tabular}{ll}
\hline
\textbf{Hyperparameter} & \textbf{Tested Values (Optimal in Bold)} \\
\hline
Learning Rate (lr)      & \textbf{4.53e-5} (from log-uniform range $10^{-5}$ to $10^{-3}$) \\
Optimizer               & AdamW, \textbf{Adam} \\
Weight Decay            & \textbf{1.1e-6} (from log-uniform range $10^{-6}$ to $10^{-4}$) \\
Batch Size              & 16, 32, \textbf{64} \\
\hline
\end{tabular}
\end{table}

\begin{table}[h!]
\centering
\caption{SSMIS hyperparameter tuning search space and the selected optimal values (in bold).}
\label{tab:hyperparams_SSMIS}
\renewcommand{\arraystretch}{1.3} 
\begin{tabular}{ll}
\hline
\textbf{Hyperparameter} & \textbf{Tested Values (Optimal in Bold)} \\
\hline
Learning Rate (lr)      & \textbf{2.104e-5} (from log-uniform range $10^{-5}$ to $10^{-3}$) \\
Optimizer               &  AdamW , \textbf{Adam}\\
Weight Decay            & \textbf{4.76e-5} (from log-uniform range $10^{-6}$ to $10^{-4}$) \\
Batch Size              &16, 32, \textbf{64} \\
\hline
\end{tabular}
\end{table}

\subsection{Framework Performance on SSMI}

\begin{figure*}[htbp]
\captionsetup[subfigure]{font=footnotesize}
    \centering
    \begin{subfigure}{\textwidth}
        \centering
        \includegraphics[width=\linewidth]{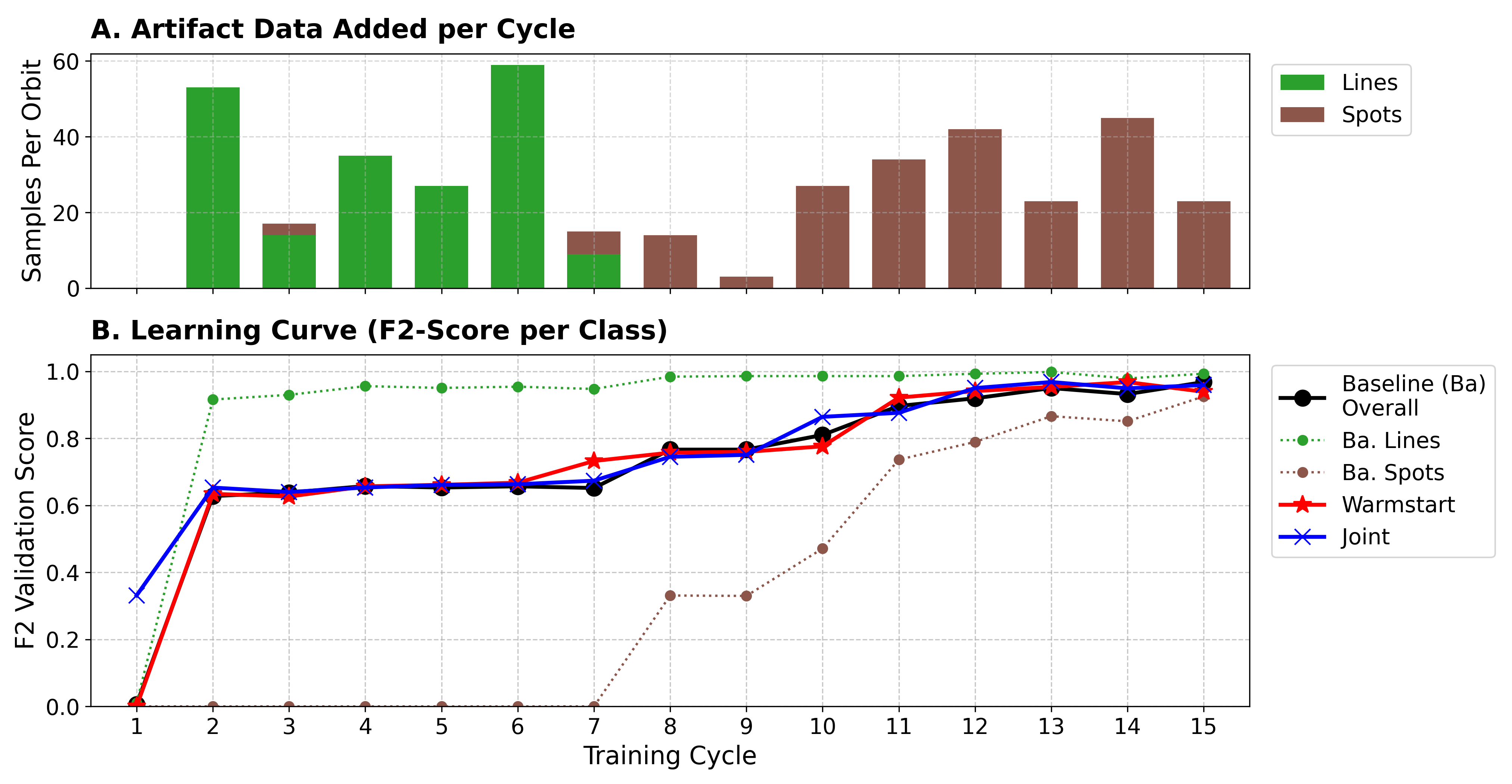}
        \caption{Artifact detection model performance on the SSMI sensor data. Top: Marginal artifact samples introduced per cycle. Bottom: Evolution of $F_{2}$-scores for the Baseline, Warm-Start, and Joint models.}
        \label{fig:SSMIperf}
    \end{subfigure}
    
    \vspace{1em} 
    
    \begin{subfigure}{\textwidth}
        \centering
        \includegraphics[width=\linewidth]{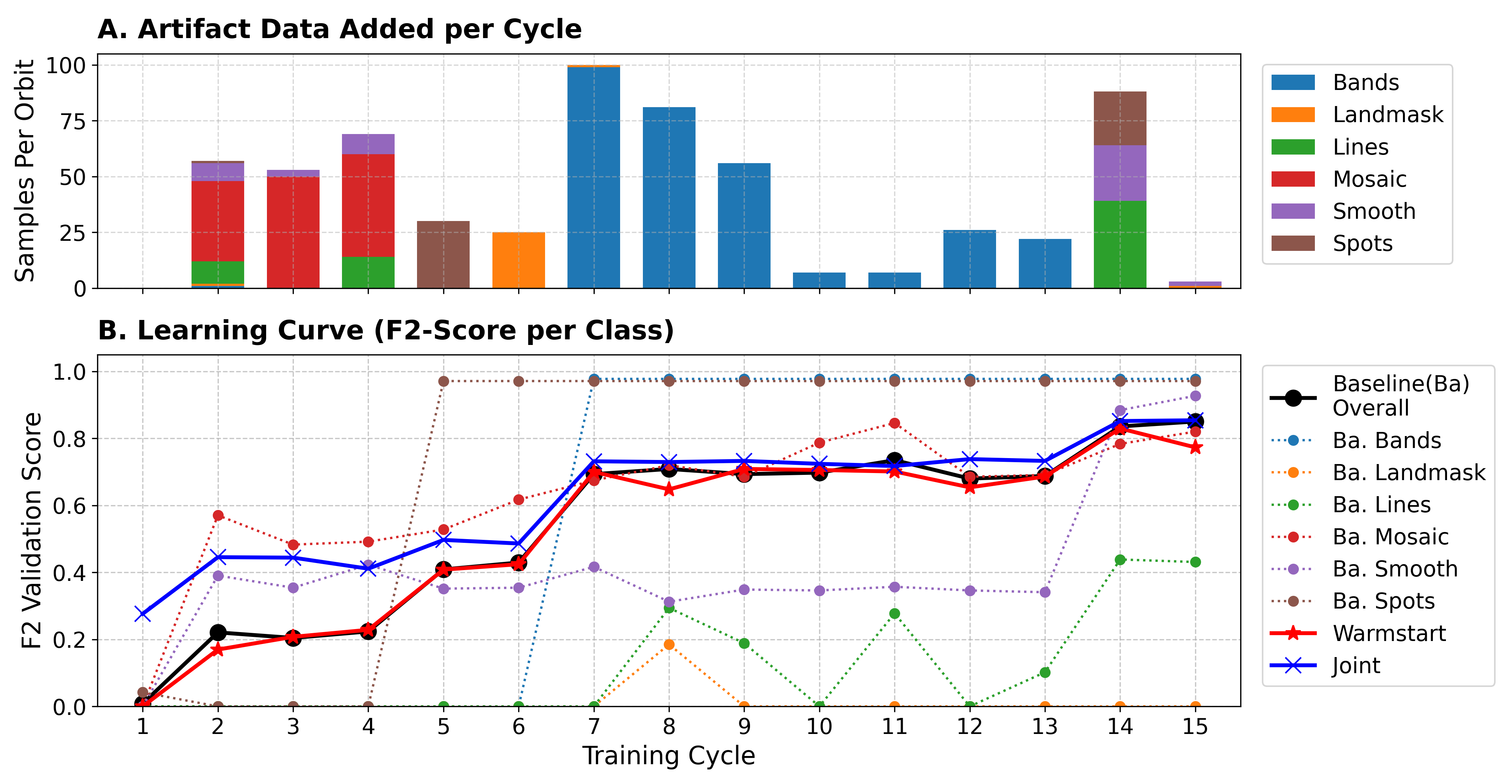}
        \caption{Artifact detection model performance on the SSMIS sensor data. Top: Marginal artifact samples introduced per cycle. Bottom: Evolution of $F_{2}$-scores for the Baseline, Warm-Start, and Joint models.}
        \label{fig:SSMISperf}
    \end{subfigure}
    
    \caption{Comparative performance evolution across the operational lifecycles of the (a) SSMI and (b) SSMIS sensors. The upper sub-panels (A) display the marginal volume of specific artifact classes introduced at each sequential training cycle. The lower sub-panels (B) track the resulting $F_{2}$-scores evaluated on an independent hold-out validation set, allowing for a direct comparison of the Baseline (Target-Only), Warm-Start (Sequential Transfer), and Joint (Data Fusion) training paradigms under varying degrees of class-specific data scarcity.}
    \label{fig:performance}
\end{figure*}

Figure \ref{fig:SSMIperf} illustrates the performance assessment for the SSMI sensor. The upper sub-panel details the data composition of each sequential cycle, while the lower sub-panel tracks the corresponding $F_2$ scores for the Baseline, Warm-Start, and Joint models.

Focusing first on the Baseline models, we observed that the overall $F_2$ score consistently increased as the volume of available data increased, though the rate of improvement varied significantly across artifact classes. For "Lines" artifacts, a performance exceeding 0.9 was achieved following the first orbit containing 53 samples; subsequent improvements for this class were consistent but marginal. In contrast, the detection capacity for "Spots" only began to increase after the introduction of an orbit with significant spot samples at Cycle 8. A comparable performance  ($F_{2} > 0.9$) for this class was not reached until Cycle 14, following the inclusion of eight distinct orbits containing exclusively spot samples.

The Joint model mitigated data scarcity through cross-sensor fusion. In Cycle 1, due to a lack of SSMI artifacts, the Baseline and Warm-Start models failed to detect any anomalies in the validation set. By leveraging the SSMIS repository, the Joint model achieved an initial $F_2$ score of 0.33. Additionally, the Joint model required the fewest epochs to converge across all three paradigms. As SSMI artifact samples arrive in cycle 2, its performance aligned with that of the isolated models, achieving comparable $F_2$ scores across the remaining cycles. 

\subsection{Framework Performance on SSMIS}

As shown in Figure \ref{fig:SSMISperf}, the model's performance for the SSMIS sensor was dictated by the arrival of specific artifact classes. The early influx of "Mosaic" samples in Cycle 2 quickly established a functional detection capacity ($F_2 \approx 0.57$) for this class. Subsequent surges in "Spots" (Cycle 5) and "Bands" (Cycle 7) triggered immediate performance spikes for those anomalies ($F_2 > 0.97$), elevating the overall model performance. Artifacts like "Smooth" arrived in smaller amounts in each cycle and required significantly more cumulative data, reaching reliable detection levels ($F_2 \approx 0.88$) only by Cycle 14, when four different orbits containing some "Smooth" classes arrived.

The Warm-Start model exhibited an overall learning trajectory highly comparable to the Baseline, demonstrating that cross-sensor pre-training alone could not overcome the absence of class-specific data in early cycles. Conversely, the Joint model successfully mitigated this early data scarcity through data fusion. In Cycle 1, with zero SSMIS artifacts available, the Joint model achieved an initial $F_2$ score of 0.28, whereas both the Baseline and Warm-Start models failed to detect any anomalies. The Joint model maintained a significant performance advantage ($F_2 \approx 0.44$ vs $0.22$) across Cycles 2 through 4. As the volume of target SSMIS artifacts increased from Cycle 7 onward, the Joint model's performance aligned with the Baseline and Warm-Start, with all three paradigms converging to a final $F_2$ score of approximately 0.85 by Cycle 15.

\subsection{Orbit-Level Classification Assessment}

The orbit-level classification performance was evaluated against a validation set of whole orbits, 6 positive and 35 negative orbits for SSMI, and 5 positive and 35 negative orbits for SSMIS, using the chronologically held-out orbits described in Section \ref{step:4}. Tables \ref{tab:orbit_metrics_ssmi} and \ref{tab:orbit_metrics_ssmis} report the per-cycle metrics for both the Baseline (Target-Only) and the Joint (Data Fusion) paradigms, alongside the SPEEDe algorithm~\cite{tanAutomatedQualityControl2024} as an external benchmark.

Considering first the Baseline paradigm, both the proposed framework and SPEEDe demonstrate high proficiency in distinguishing artifact-contaminated from regular orbits once sufficient target data has accrued. For SSMIS, the Baseline achieved ROC-AUC and PR-AUC of 1.000 as early as Cycle 2. For SSMI, performance was consistently high after Cycle 2, with ROC-AUC and PR-AUC values generally exceeding 0.9, eventually achieving 1.000 by Cycle 13. The SPEEDe framework achieved high, though not perfect, scores across both sensors (ROC-AUC of 0.995 for SSMI and 0.960 for SSMIS). 

Critically, the Baseline paradigm shows weak artifact detection in the first cycle (ROC-AUC of 0.586 for SSMI and 0.320 for SSMIS), as no target-sensor artifacts have yet been observed. However, the Joint (Data Fusion) paradigm successfully mitigates this cold-start limitation; by leveraging cross-sensor data, the Joint model achieves a Cycle 1 ROC-AUC of 0.862 for SSMI and 0.989 for SSMIS, proving its operational value immediately upon deployment.

For both satellites, the Brier Score exhibits a consistent downward trend, decreasing from 0.138 to 0.021 for SSMI and from 0.106 to 0.039 for SSMIS. Notably, this improvement in performance continues even in cycles where the AUC scores remain stable or fluctuate slightly.
 
\begin{table}[h!]
\centering
\caption{Per-cycle orbit-level evaluation metrics for the Baseline (Target-Only) and Joint (Data Fusion) paradigms on the held-out whole-orbit validation set for SSMI. The final row reports the SPEEDe benchmark; its Brier Score is not applicable as SPEEDe outputs a reconstruction-error score rather than a probability.}
\label{tab:orbit_metrics_ssmi}
\renewcommand{\arraystretch}{1.25}
\setlength{\tabcolsep}{5pt}
\footnotesize
\begin{tabular}{lccc|ccc}
\hline
Paradigm & \multicolumn{3}{c}{Baseline} & \multicolumn{3}{c}{Joint} \\
Metric & \begin{tabular}{@{}c@{}}ROC-\\AUC\end{tabular} & \begin{tabular}{@{}c@{}}PR-\\AUC\end{tabular} & \begin{tabular}{@{}c@{}}Brier\\Score\end{tabular} & \begin{tabular}{@{}c@{}}ROC-\\AUC\end{tabular} & \begin{tabular}{@{}c@{}}PR-\\AUC\end{tabular} & \begin{tabular}{@{}c@{}}Brier\\Score\end{tabular} \\
\hline
Cycle 1 & 0.586 & 0.333 & 0.138 & 0.862 & 0.674 & 0.047 \\
Cycle 2 & 0.976 & 0.924 & 0.042 & 0.933 & 0.883 & 0.037 \\
Cycle 3 & 0.967 & 0.910 & 0.041 & 0.971 & 0.917 & 0.037 \\
Cycle 4 & 0.929 & 0.881 & 0.042 & 0.986 & 0.944 & 0.038 \\
Cycle 5 & 0.929 & 0.881 & 0.041 & 0.924 & 0.879 & 0.038 \\
Cycle 6 & 0.990 & 0.948 & 0.042 & 0.981 & 0.933 & 0.038 \\
Cycle 7 & 0.952 & 0.896 & 0.042 & 0.948 & 0.892 & 0.040 \\
Cycle 8 & 0.914 & 0.764 & 0.031 & 0.910 & 0.789 & 0.032 \\
Cycle 9 & 0.924 & 0.802 & 0.030 & 0.848 & 0.625 & 0.031 \\
Cycle 10 & 0.976 & 0.924 & 0.025 & 0.957 & 0.850 & 0.025 \\
Cycle 11 & 0.986 & 0.944 & 0.024 & 0.938 & 0.833 & 0.026 \\
Cycle 12 & 0.981 & 0.917 & 0.026 & 0.981 & 0.933 & 0.025 \\
Cycle 13 & 1.000 & 1.000 & 0.023 & 0.952 & 0.896 & 0.023 \\
Cycle 14 & 1.000 & 1.000 & 0.024 & 0.986 & 0.944 & 0.024 \\
Cycle 15 & 1.000 & 1.000 & 0.021 & 0.976 & 0.924 & 0.023 \\
SPEEDe & 0.995 & 0.976 & --- & 0.995 & 0.976 & --- \\
\hline
\end{tabular}
\end{table}


\begin{table}[h!]
\centering
\caption{Per-cycle orbit-level evaluation metrics for the Baseline (Target-Only) and Joint (Data Fusion) paradigms on the held-out whole-orbit validation set for SSMIS. The final row reports the SPEEDe benchmark; its Brier Score is not applicable as SPEEDe outputs a reconstruction-error score rather than a  probability.}
\label{tab:orbit_metrics_ssmis}
\renewcommand{\arraystretch}{1.25}
\setlength{\tabcolsep}{5pt}
\footnotesize
\begin{tabular}{lccc|ccc}
\hline
Paradigm & \multicolumn{3}{c}{Baseline} & \multicolumn{3}{c}{Joint} \\
Metric & \begin{tabular}{@{}c@{}}ROC-\\AUC\end{tabular} & \begin{tabular}{@{}c@{}}PR-\\AUC\end{tabular} & \begin{tabular}{@{}c@{}}Brier\\Score\end{tabular} & \begin{tabular}{@{}c@{}}ROC-\\AUC\end{tabular} & \begin{tabular}{@{}c@{}}PR-\\AUC\end{tabular} & \begin{tabular}{@{}c@{}}Brier\\Score\end{tabular} \\
\hline
Cycle 1 & 0.320 & 0.283 & 0.106 & 0.989 & 0.927 & 0.041 \\
Cycle 2 & 1.000 & 1.000 & 0.041 & 0.966 & 0.796 & 0.041 \\
Cycle 3 & 0.971 & 0.900 & 0.041 & 0.966 & 0.796 & 0.041 \\
Cycle 4 & 0.983 & 0.925 & 0.041 & 0.966 & 0.796 & 0.041 \\
Cycle 5 & 0.994 & 0.967 & 0.041 & 0.971 & 0.826 & 0.041 \\
Cycle 6 & 1.000 & 1.000 & 0.040 & 0.983 & 0.903 & 0.040 \\
Cycle 7 & 1.000 & 1.000 & 0.041 & 0.977 & 0.885 & 0.041 \\
Cycle 8 & 1.000 & 1.000 & 0.040 & 0.983 & 0.925 & 0.040 \\
Cycle 9 & 1.000 & 1.000 & 0.040 & 0.983 & 0.925 & 0.040 \\
Cycle 10 & 1.000 & 1.000 & 0.041 & 0.983 & 0.925 & 0.040 \\
Cycle 11 & 1.000 & 1.000 & 0.040 & 0.983 & 0.925 & 0.040 \\
Cycle 12 & 1.000 & 1.000 & 0.041 & 0.977 & 0.885 & 0.040 \\
Cycle 13 & 1.000 & 1.000 & 0.041 & 0.989 & 0.943 & 0.041 \\
Cycle 14 & 1.000 & 1.000 & 0.040 & 0.983 & 0.925 & 0.039 \\
Cycle 15 & 0.989 & 0.927 & 0.039 & 0.977 & 0.911 & 0.040 \\
SPEEDe & 0.960 & 0.883 & --- & 0.960 & 0.883 & --- \\
\hline
\end{tabular}
\end{table}

\subsection{Spatial Distribution and Probabilistic Mapping of Artifacts}

\begin{figure*}
    \centering
    \includegraphics[width=\textwidth]{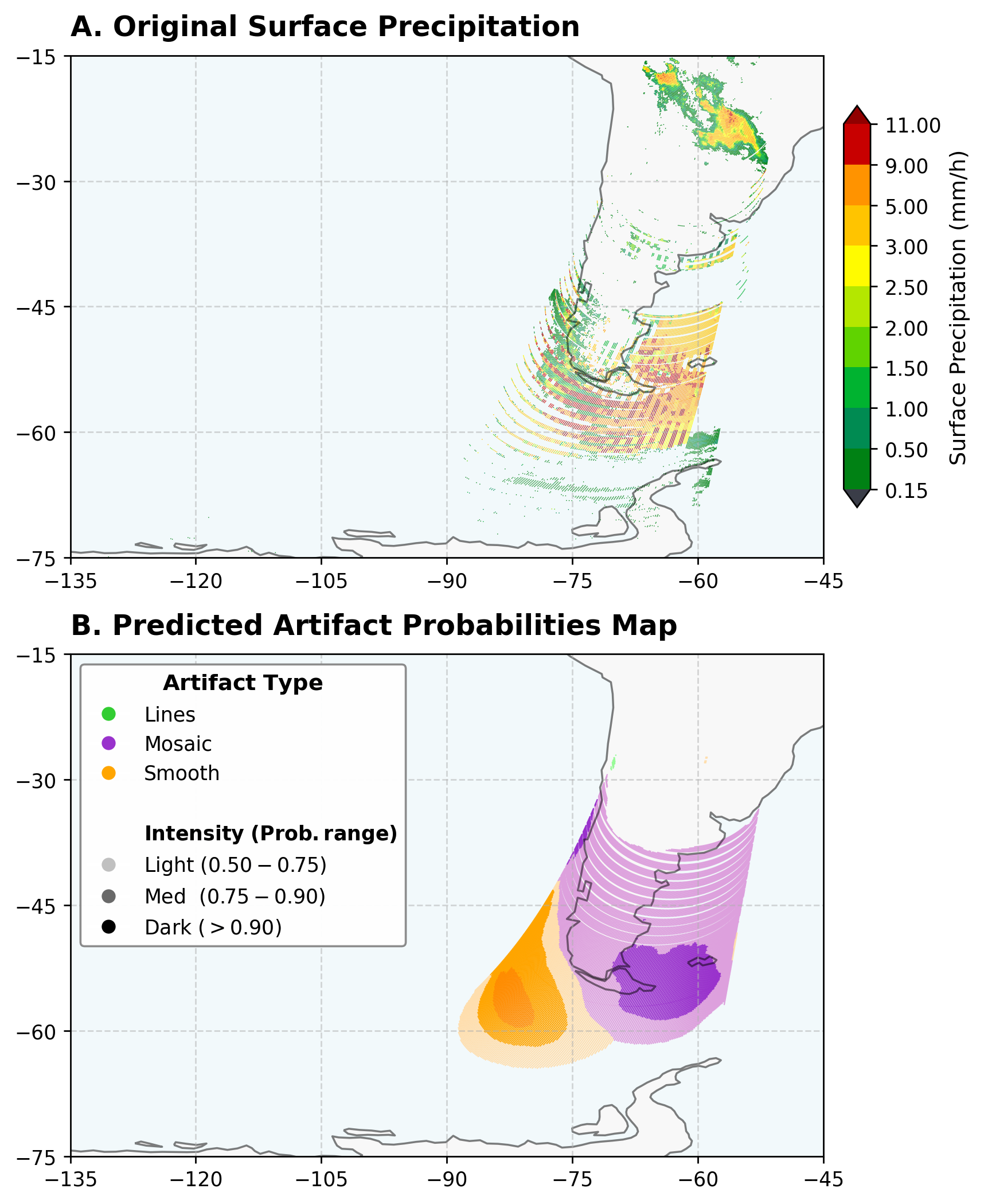}
    \caption{(A) An example SSMIS anomalous orbit  with Mosaic and Smooth artifacts in southern South America (B) Corresponding map displaying the framework's pixel-level artifact probabilities across the orbit domain (valid rain pixels only). A transparency threshold is applied where $p < 0.5$ is invisible and higher probabilities are shown in gradient. Darker color indicates higher model confidence}
    \label{fig:mapa}
\end{figure*}

As shown in Figure \ref{fig:mapa}, the framework allows to assign an artifact probability to each valid pixel across the orbit domain (no-rain pixels are not considered valid). These probabilities are visualized using a transparency and color scheme: regions with low confidence ($probabilities<0.5$) remain transparent, while areas of increasing probability are highlighted in progressively deeper shades of color, providing an actionable tool for human reviewers to interpret and locate the potential artifacts.

\subsection{Minimum Description Length per Artifact Class}

A significant operational advantage of the proposed framework is its capacity for iterative refinement through an active feedback loop. When the model generates a false positive or a false negative, these specific samples can be immediately added to the training pool to refine the decision boundary in subsequent cycles.

This capacity for iterative refinement is particularly vital when addressing the morphological disparities between different satellite sensors. By analyzing how each individual, randomly sampled orbit sub-window alters the model's ability to identify validation samples of the same class, we observe that the threshold for sufficient data, the minimum description length required to learn a class successfully, is not a universal number. Instead, it is highly dependent on sensor-specific textures and the spatial density of the anomaly.

\begin{figure*}
    \centering
    \begin{subfigure}{0.9\textwidth}
        \centering
        \includegraphics[width=\linewidth]{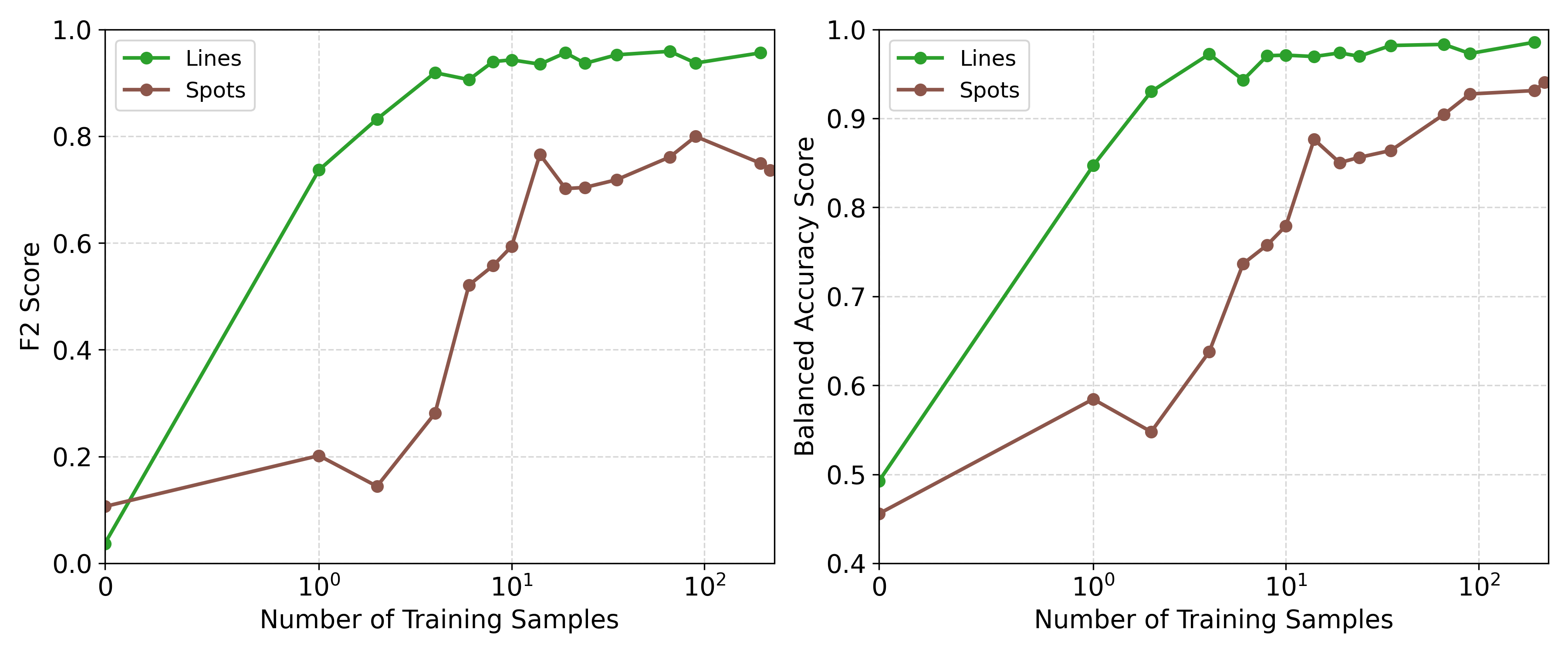}
        \caption{}
        \label{fig:SSMI_samples}
    \end{subfigure}
    
    \vspace{1em} 
    
    \begin{subfigure}{0.9\textwidth}
        \centering
        \includegraphics[width=\linewidth]{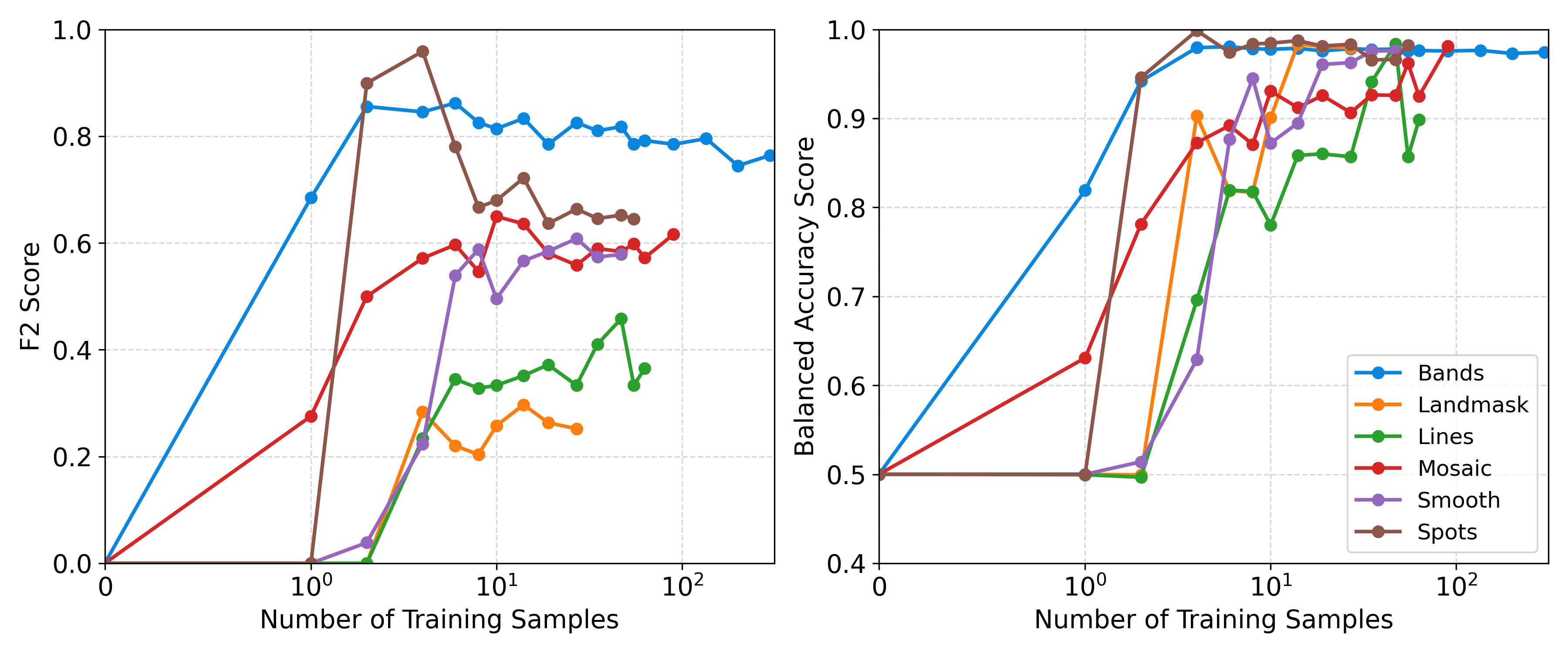}
        \caption{}
        \label{fig:SSMIS_samples}
    \end{subfigure}
    
    \caption{Impact of incremental sample introduction on class-specific detection performance. The plot illustrates the sensitivity of the $F_2$ score (left panels) and to the $BA$ (right panels) to the addition of individual, randomly sampled orbit sub-windows for (a) SSMI and (b) SSMIS.}
    \label{fig:per_sample_perf}
\end{figure*}

Although artifact classes such as "Lines" or "Spots" share the same nomenclature across sensors, their physical characteristics dictate different learning trajectories. In SSMI, for instance, "Lines" frequently appear as densely clustered features. This spatial density results in an extremely short description length; as seen in Figure \ref{fig:per_sample_perf}, the model achieves an F2 score of 0.919 and a $BA$ of 0.972 after exposure to just 4 samples. SSMI "Spots," being less structured, require approximately 14 samples to cross an F2 threshold of 0.750.

Conversely, artifacts in SSMIS tend to manifest as more isolated anomalies, altering the required data volume for convergence. The disparity is most evident in the SSMIS "Lines" class. Unlike its SSMI counterpart, the model requires nearly 40 samples to surpass an F2 score of 0.400, empirically demonstrating that sparser spatial features demand a significantly longer description length to establish a reliable decision boundary.

Intra-sensor variance in SSMIS further highlights this dependency on structural complexity. Highly structured, rigid features like "Bands" are learned almost instantaneously, reaching an F2 of 0.856 with merely 2 samples. In contrast, anomalies contingent on external geographical contexts, such as the "Landmask" class, exhibit the highest mathematical complexity. The model struggles to isolate this specific signature, failing to exceed an F2 score of 0.300 even after exposure to 27 samples, suggesting that boundary-dependent artifacts require substantially larger training volumes to be effectively mapped.

\subsection{Code and Data Repository}

The complete artifact-detection framework; trained models, processing workflow,
and analysis code, are available through the project GitHub repository \url{https://github.com/afmonsalves/SAPAD.git}, which provides step-by-step scripts for orbit preprocessing, patch generation, data splitting, model training, cross-validation, and evaluation, together with the Baseline, Warm-Start, and Joint strategies best-models. The labeled and pre-augmented set of 70$\times$70 artifact and regular patches for SSMI and SSMIS, is published separately as a citable Zenodo deposit (\url{https://doi.org/10.5281/zenodo.21793955}), while the \texttt{orbit\_lists/} folder in the repository enumerates the original, unprocessed whole-orbit granules (from NASA GES DISC) from which those patches were derived.

\section{Discussion}\label{chap:discussion}

In this study, a supervised, cross-sensor anomaly-detection framework was successfully developed and validated, capable of identifying and classifying specific passive microwave precipitation retrieval artifacts in data-scarce environments.

\subsection{Novelty and Operational Advantages}\label{ch:novelty}

A notable finding is the divergence between high binary classification scores, such as the ROC-AUC and PR-AUC metrics in Tables \ref{tab:orbit_metrics_ssmi} and \ref{tab:orbit_metrics_ssmis}, and the more conservative $F_2$ scores observed during the incremental learning cycles in Figure \ref{fig:performance}. This disparity reveals a fundamental characteristic of the framework: while the model at early cycles may occasionally misclassify one specific artifact type as another, it remains exceptionally proficient at the primary task of distinguishing valid precipitation orbit from an anomalous one. The optimization of the model is intentionally skewed toward a "cautious" operational posture. In a meteorological context, the cost of allowing a contaminated orbit to pass (a False Negative) is significantly higher than the cost of flagging a regular orbit for further review (a False Positive). Consequently, the high binary scores in Tables \ref{tab:orbit_metrics_ssmi} and \ref{tab:orbit_metrics_ssmis} reflect the model's robustness in ensuring data integrity, even when fine-grained class differentiation is still evolving.

The consistent improvement of the Brier Score further supports this interpretation of model maturity. As the framework processes a greater volume of orbits, it becomes increasingly well-calibrated, even if the ability to separate good vs anomalous orbits marginally improves (e.g. SSMIS ability to separate good from bad orbits excels first at cycle 7). This reduction in the Brier Score indicates that the probabilistic assignments are becoming more representative of the actual likelihood of an artifact's presence.

A key distinction between our framework and SPEEDe lies in the nature of their outputs. While SPEEDe operates as a non-interpretable classifier, where the score is based on the difference between the pixel values of the orignal sample and a reconstructed image, our proposed framework provides an independent probability for each possible artifact type.

\begin{figure}
    \centering
    \includegraphics[width=\linewidth]{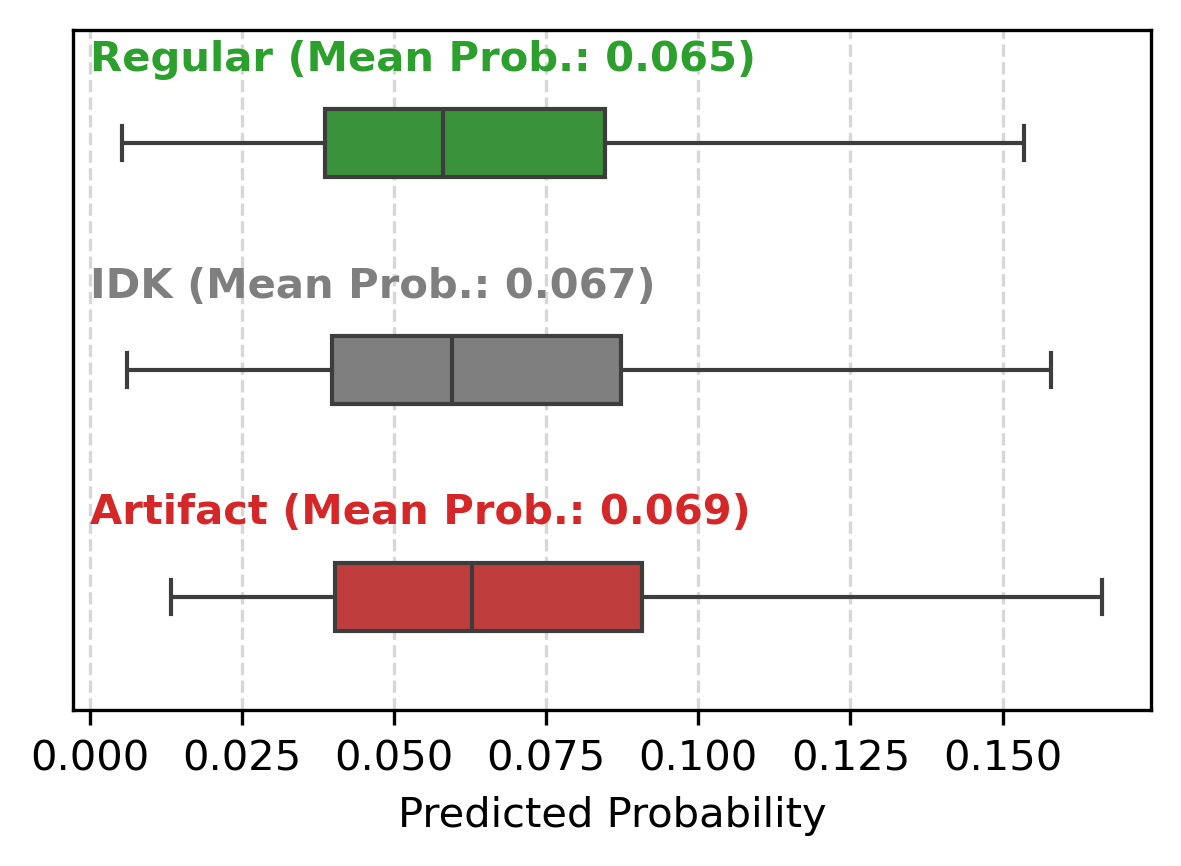}
    \caption{Predicted probability distributions by class category. Boxplots illustrate the interquartile range of model predictions for regular, artifact, and "IDK" validation groups. The IDK samples occupy the transition zone between nominal precipitation and confirmed artifacts, quantitatively validating the model's ability to mirror expert uncertainty in ambiguous meteorological scenes.}
    \label{fig:boxplot}
\end{figure}

The manual labeling process revealed a significant subset of data where the criteria for artifact identification remained unclear. Because an absolute ground truth did not exist for these scenes, we introduced an auxiliary 'IDK' (indeterminate) category to isolate windows with conflicting signals that an expert could not definitively classify. Excluding these samples from the training process allowed us to prioritize high-confidence data over noise. A supervised model should not be tasked with classifying features that exceed the interpretive limits of a human expert. The interquartile range of probability predictions for regular, artifact, and IDK validation scenes is shown in Figure \ref{fig:boxplot}. Notably, the predictions for the IDK group reside in the transition zone between nominal precipitation and identified artifacts. This confirms that even without explicit training on these features, the model recognizes levels of irregularity that differentiate them from standard orbits. This outcome validates our decision to isolate these samples, as the model’s probabilistic uncertainty effectively replicates the thought process of the authors used during the labeling phase.

\subsection{Methodological Limitations}\label{ch:limitations}

Data scarcity and classification ambiguity posed significant challenges in this study. The precision of labeling was more determinant than the total volume of data, as even a few mislabeled samples could disrupt model convergence. Over multiple labeling iterations for both sensors, it became clear that features originally viewed as single classes often possessed multi-class characteristics. Precisely defining the visual attributes that characterize a class is therefore critical to this framework. Ultimately, attempting to force ambiguous data into broad categories proved detrimental to the model’s discriminative performance.

An important limitation of the Joint and Warm-Start paradigms is their reliance on the geometric compatibility of the aggregated data. Because both SSMI and SSMIS operate as conical scanners, they share fundamentally similar field-of-view characteristics. The efficacy of feature transfer or joint training across divergent scanning geometries, such as adapting from a conical to a cross-track scanner, remains untested. In this case, the isolated Baseline paradigm remains the most structurally robust approach, as it is self-contained and natively learns to map only the specific distortions and artifact morphologies inherent to its own instrument.

\subsection{Recommended Next Steps}\label{ch:whatsnext}

While this study demonstrated the versatility of the MobileNetV3-Large backbone, quantifying the exact number of training samples required across heterogeneous satellite platforms remains a fundamental challenge in scarce-data environments. Because visually similar artifacts require different exposure volumes due to sensor-specific textures, establishing a universal data threshold for operational readiness is currently impractical. Future research must investigate how underlying feature-extraction requirements scale across a broader array of passive microwave sensors.

Also, exploring the "IDK" class offers a highly promising avenue for future research through the integration of human-in-the-loop (HITL) validation frameworks. Because these indeterminate samples occupy the boundary between nominal precipitation and confirmed artifacts, they represent a critical knowledge gap that current architectures cannot autonomously resolve. Future efforts should focus on deploying distributed classification interfaces, such as web-based platforms for evaluating anomalous orbit screenshots, where domain experts can anonymously assess and categorize these conflicting signals to serve as a global benchmark for satellite precipitation anomalies.

\section{Conclusion}\label{chap:conclusion}

This study addresses the challenge of ensuring data integrity in near-real-time satellite precipitation products, where ground truth is often unavailable. For many end-users, distinguishing natural rain patterns from sensor-specific artifacts remains difficult without specialized training, leading to potential misinterpretations of surface precipitation characteristics. To alleviate this problem, we developed a sensor-agnostic framework that can identify anomalies across disparate satellite platforms from their first day of operation. This framework provides interpretable, explainable diagnostic output detailing the location, classification, and probabilistic certainty of detected artifacts. It demonstrates robust capacity to improve detection sensitivity as new artifact categories are introduced through an iterative training cycle.

The developed methodology establishes a Baseline model development process that is inherently sensor-agnostic, enabling rapid adaptation to new satellite platforms with minimal computational overhead. This flexibility is particularly critical given the industry shift toward SmallSat constellations, which introduce dozens or hundreds of new data sources. The framework’s ability to provide on-the-go quality assessment ensures that these emerging high-volume data streams remain reliable and operationally viable from deployment. Furthermore, when historical anomaly data from analogous platforms is accessible, the Joint training methodology provides a powerful alternative. By aggregating cross-sensor artifact datasets, the Joint model leverages generalized features to directly mitigate data starvation in early operational cycles, when no artifact information is yet available for the target sensor. This multi-sensor fusion accelerates mathematical convergence and enhances classification performance for otherwise sparse artifact classes.

\section*{Acknowledgments}

This research was supported by the National Aeronautics and Space Administration (NASA) Precipitation Measurement Mission (PMM) program under grant no. 80NSSC22K0604. H. Moreno was supported by the National Oceanic and Atmospheric Administration – Cooperative Science Center for Earth System Sciences and Remote Sensing Technologies under the Cooperative Agreement Grant \# NA22SEC4810016. The authors would like to express their gratitude to Jackson Tan and Yi Song at NASA and to Paula Brown at the Cooperative Institute for Research in the Atmosphere (CIRA) for providing the artifact datasets and for their support throughout this study. The statements, findings, conclusions, and recommendations are those of the author(s) and do not necessarily reflect the views of NASA or NOAA.

\section*{AI Disclaimer}

Generative AI tools were used solely for code optimization and for formatting tables and plots. The authors maintained full control over the refinement of these outputs. All scientific analysis, conclusions, and the final narrative were authored exclusively by the researchers, who remain fully responsible for the content and integrity of this article.

\bibliographystyle{unsrt}
\bibliography{manual_bib, references}

\vfill

\end{document}